%% file: guardauto.tex
\documentclass[journal,twocolumns]{IEEEtran}
\usepackage{amsfonts}
\usepackage{amsmath}
\usepackage{amssymb}
\usepackage{slashbox}
\usepackage{graphicx}
\usepackage{epstopdf}
\usepackage{graphicx}
\usepackage{textcomp}
\usepackage{subfigure}
\usepackage[perpage]{footmisc}
\usepackage{makecell}
\usepackage[section]{algorithm}
\usepackage{algpseudocode}
\usepackage{xcolor}
\usepackage{listings}
\usepackage{xspace}
\usepackage{enumitem}
\usepackage{pbox}
\usepackage{cite}
\usepackage{hyperref}
\usepackage{breakurl}

\newcommand{\subparagraph}{}
\usepackage{titlesec}

\newtheorem{ex}{Example}[section]
\newtheorem{definition}{Definition}[section]
\newcommand{\tool}{Guardauto\xspace}

\newcommand{\review}[2]{{#2}}

\renewcommand{\ref}{\hyperref}

\ifCLASSINFOpdf
\else
\fi

\begin{document}
%
\title{Guardauto: A Decentralized Runtime Protection System for Autonomous Driving}

\author{
	Kun~Cheng,~Yuan~Zhou,~Bihuan~Chen,
		Rui~Wang,~Yuebin~Bai~and~Yang~Liu
\thanks{K.~Cheng, R.~Wang~and~Y.~Bai are with School of Computer Science and Engineering, Beihang University, Beijing, China 100191. Email: \{chengkun, wangrui, byb\}@buaa.edu.cn. R.~Wang is the corresponding author.}
\thanks{Y.~Zhou and Y.~Liu are with School of Computer Science and Engineering, Nanyang Technological University, Singapore 679398. Emails: \{y.zhou, yangliu\}@ntu.edu.sg}%
\thanks{B.~Chen is with School of Computer Science, Fudan University, Shanghai, China 201203. Email: bhchen@fudan.edu.cn }
}

\maketitle

%

\input{SourceCode/abstract}


%
\IEEEpeerreviewmaketitle

\input{SourceCode/introduction}
\input{SourceCode/relatedwork}
\input{SourceCode/motivation}
\input{SourceCode/decouple_model}
\input{SourceCode/adaptation_design}
\input{SourceCode/implementation}

\input{SourceCode/evaluation_tii}

\input{SourceCode/conclusion}
\bibliographystyle{IEEEtran}
\bibliography{SourceCode/reference}
\ifCLASSOPTIONcaptionsoff
  \newpage
\fi

\end{document}

%% file: SourceCode/abstract.tex

\begin{abstract}	
	Due to the broad attack surface and the lack of runtime protection, potential safety and security threats hinder the real-life adoption of autonomous vehicles.
	Although efforts have been made to mitigate some specific attacks, there are few works on the protection of the self-driving system.
	This paper presents a decentralized self-protection framework called \tool to protect the self-driving system against runtime threats.
	First, \tool proposes an isolation model to decouple the self-driving system and isolate its components~with a set of partitions.
	Second, \tool~provides self-protection mechanisms for each target component, which combines different methods to monitor the target execution and plan adaption actions accordingly.
	Third, \tool \ provides cooperation among local self-protection mechanisms to identify the root-cause component in the case of cascading failures affecting multiple components.
	A prototype has been implemented and evaluated on the open-source autonomous driving system Autoware.
	Results show that~\tool could effectively mitigate runtime failures and attacks, and protect the control system with acceptable performance overhead.
\end{abstract}

%% file: SourceCode/introduction.tex
\section{Introduction}\label{intro}

Since the last decade, autonomous vehicles 
have been attracting more and more attention.
In particular, since Google announced~its self-driving project in 2010, many companies and institutes have been devoting themselves into this domain, 
such as Waymo, Tesla Autopilot, and Baidu Apollo Open Platform~\footnotemark.

\begin{figure}[!t]
	\centering
	\includegraphics[width=0.75\linewidth]{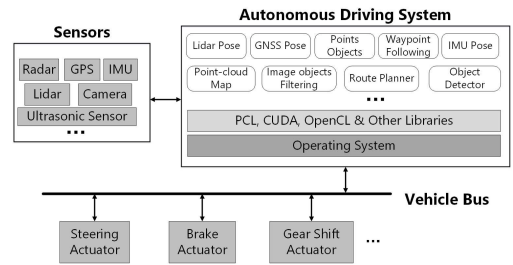}
	\caption{\review{5-1}{A comprehensive autonomous vehicle system.}}
	\label{avsystem}
\end{figure}

Autonomous vehicles are complex cyber-physical systems.
As shown in Fig. \ref{avsystem}, the hardware of such a system includes sensors, electronic control units (ECUs) or actuators and other embedded devices.
The control software system, i.e., the software running in a control computer, manages the operation of different hardware devices, which is narrowly referred to as the self-driving system or the autonomous driving systems (ADS) in this paper.
Specifically, the ADS collects data from sensors to complete tasks such as path planning and navigation control, and sends commands to ECUs via vehicle bus, e.g., Controller Area Network (CAN), to guide actuators.
In that sense, the ADS serves as the brain of a self-driving car.

Due to the complexity nature, self-driving cars have a broad~attack surface \cite{jonathan,miller2014survey} that may allow adversaries to intrude remotely~\cite{greenberg2015hackers}~(e.g., via Bluetooth, remote keyless entry, or Uconnect),
or to disrupt data links (e.g., conducting a camera, Lidar or GPS spoofing~\cite{petit2015remote,gpsAttack} to provide manipulated data).
Besides, the current self-driving system consists of dozens of various functionalities, which may spawn hundreds of user processes interacting with each other through tremendous data exchange.
As those functions are running together with other applications on top of the same kernel, any software failures or malware~\cite{vassallo2018spatially} could easily compromise the whole system or interfere the self-driving control.
Moreover, it is also challenging to track the interaction among the driving functionalities, which makes it difficult to analyze and model the exact system behavior.
Although several countermeasures
have been proposed as concluded in Sec.~\ref{related}, most of them aim at the protection of specific devices or communications against attacks, while few focuses on the protection of the ADS.
Potential attacks to such a system, e.g., disrupting critical services or blocking the normal transmission, can lead to catastrophes such as vehicles crashing and human casualties.
Thus, it is critical to keep potential intrusion or software failures from disrupting the ADS to ensure its proper execution~\cite{Schoitsch2016}.




Fault tolerance is important to protect such complicated systems.
Specifically, the protection mechanism is required to identify the abnormal execution behavior, keep the ongoing errors/failures from further jeopardizing other functionalities, and restore the system back to normal.
Except for the above functional requirements, a protection system should also be friendly to deployment and upgrade, which requires a flexible design.
Finally, the protection systems should also be resilient to failures, which is seldom considered in existing works.
Therefore, challenges arise in developing such protection scheme as the following.
\footnotetext{{Waymo, \url{https://waymo.com}. Autopilot, \url{https://www.tesla.com/autopilot}. Baidu Apollo, \url{http://apollo.auto}}}
\begin{itemize}[leftmargin=*,topsep=0pt,partopsep=0pt]
	\item  How can the protection system identify an abnormal execution.
	\item  How can it protect the rest system from a failed component.
	\item  How an the system maintain the flexibility and scalability.
	\item  How to ensure a reliable system design.
\end{itemize}
However, as traditional architecture-based solutions adopt the central adaptation management~\cite{Tziakouris-2018}, the adaptation layer may suffer from single-point failure if any of its own functionalities fail, and the mechanism lacks adequate scalability for further upgrades, which will eventually make the system less effectively to deal with new threats.
Thus, they do not well address all the challenges.
In this paper, we propose \tool, the guardian of autonomous driving systems.
First, \tool~ includes a component model to decouple the ADS, and uses a hybrid solution of virtual machine and container based partitions to achieve isolation.
Second, \tool deploys a decentralized self-protection scheme for all components,
which automatically detects anomalies or failures in a isolated component and adapts the ADS accordingly.
Finally,
instead of naively mitigating all failed components, \tool analyzes the data flow and leverages the global cooperation among local self-protection mechanisms to locate where the root cause is, and only restore that component, which ensures the system availability.
To illustrate the proposed approach, a prototype of \tool is evaluated on an open-source control system Autoware~\cite{autoware}, which is based on Robot Operating System (ROS)~\cite{ros09} and has been deployed in a variety of self-driving platforms, e.g., ZMP RoboCar~\cite{zmp}.

The main contribution of this work includes:
\begin{itemize}[leftmargin=*,topsep=0pt,partopsep=0pt]
\item By analyzing the autonomous driving system protection challenges, we propose a component isolation model from the functional perspective, which could be served as a guideline to other similar systems.
\item We present a decentralized self-protection system which merges the need of system-level reconfiguration and a resilient system design. 
\item We implement the self-protection system, and conduct evaluation to prove that such a system meets the design guideline and can effectively mitigate runtime threats.
\end{itemize}

The reset of paper is organized as follows.
Sec.~\ref{related} summaries the related works on protecting self-driving cars,
Sec.~\ref{motivation} presents the preliminaries of this work, and Sec.~\ref{designs} describes the detailed design of \tool,
Sec.~\ref{implementation} gives the implementation.
In Sec.~\ref{evaluations}, experiments are conducted to evaluate \tool.
We conclude this paper in Sec.~\ref{conclusion}.

%% file: SourceCode/relatedwork.tex
\section{Related Work}\label{related}

\subsubsection{Safety and Security of Autonomous Driving Vehicles.}
Lee et al.~\cite{secureAS} discussed how compile-time assurance, runtime protection, automated testing and architectural security could be used in automotive software. 
Seshia et al.~\cite{Seshia} discussed the major challenges in applying formal methods in the specification, design and verification of semi-autonomous vehicles.
Adler et al.~\cite{safetyengineering} proposed a safety engineering approach for self-driving vehicles through a safety superior and fault tree based analysis to identify and handle malfunction.
Several protection solutions have also been proposed in the architecture design. AUTOSAR~\cite{autosar} proposed an adaptive platform, which leveraged SoA framework and distributed computation, and also proposed identity, access, crypto and key management to ensure security.
Similar proposal can also be found in~\cite{futureAVsys}.

Moreover, a variety of methods were proposed to counter attacks.
For example, Jasen et al.~\cite{Jansen-2016} proposed a multi-receiver GPS spoofing and detection by checking the relative distance of all receivers.
However, such method required a specific deployment of antennas covering at least $26m^2$, which was not suitable for vehicular systems.
Similar work could also be found in~\cite{ccs2014}, where infrastructure help (such as road side mobile units) was used to detect the GPS jamming attack, and research~\cite{sensordtect} leveraged pairwise inconsistencies between sensors to detect transient attack or faults for GPS receivers.
Zhang et al.~\cite{safedrive} proposed a behavior model based approach to detect anomaly driving status.
However, such a model was built from in-vehicle sensor data offline without the consideration of environmental factors (e.g. other running vehicles and their influence on the driving behavior).
Cho and Shin~\cite{ccs2016} detected a specific DoS attack, where the error handling mechanism of the CAN bus was exploited to force the target ECUs enter the bus off status.
However, it required a strict timing synchronization to inject and launch the attack.
Authentication and encryption for in-vehicle communication were also proposed~\cite{bmw2013,attackcan}.
However, the overhead of such mechanisms
hindered the practical adoption.
Recently, Steger \emph{et al} proposed a framework for secure and dependable wireless software update on ECUs, \cite{steger2018efficient},
which adopted a strong authentication and encryption to secure ECU firmware update from data alteration or leak.
However, other possible threats, such as DoS attacks, were not considered.

The security and privacy concerns of vehicular networks (e.g., V2X) also draw a lot of interest.
Most existing works focused on providing~\cite{ALNASSER2019} 1) cryptography-based solutions, such as authentication and encryption; 2) behavior-based mechanisms for monitoring, such as rule-based monitoring and weighted-sum method; 3) identity-based solutions, such as using pseudonyms to preserve the privacy of vehicles' locations.
For example,
Wu \emph{et al}~\cite{reviewer4-2} proposed a resource management scheme to secure vehicle-to-cloud communication from eavesdropping attacks, which focused on physical layer security from the perspective of  radio resource allocation.
Sedjelmaci \emph{et al}~\cite{SEDJELMACI201774} proposed a rule-based technique to model the normal behavior of a vehicle with the help of roadside units, which was used in the intrusion detection system against false alerts and Sybil attacks.

All the above works aim at protecting the self-driving system from specific attacks, especially on the cryptography mechanism, sensor signal and communication of either internal (e.g. CAN bus) or external networks (e.g. V2X).
As far as we know, there is few on the system protection.


\review{2-2}{
\subsubsection{Self-Protecting Systems.}
Self-protecting systems are a class of self-adaptive systems that detect and mitigate security threats at runtime~\cite{Yuan-2014}.
C.-H. Lung et al.~\cite{LUNG2016311} proposed an architectural self-healing framework for concurrent environment, which instantly switched to an alternative when Half-Sync/Half-Async failures were detected.
Rodr{\'\i}guez et al. \cite{rodriguez2016model} proposed to protect systems by adding security layers when attacks are discovered.
The selection of the added security layer was based on the  evaluation of the protection mechanisms via dynamic Bayesian networks.
Frtunikj et al.~\cite{adaptiveerror} proposed an adaptive error and state management model for autonomous~cars, which aimed to allow the system to automatically adapt after part of the system failed. However, their model treated each error independently without considering the possible failure propagation. Instead, \tool deploys a cooperation mechanism to solve that.
To secure the web applications, Chen et al.~\cite{QChen-2014} developed a self-protecting system for the Internet of Things to autonomously estimate, detect and react to cyber attacks by filtering malicious packets or replacing compromised nodes.
Watanabe et al.~\cite{dac-18}  studied runtime monitoring through the signal temporal logic to detect undesirable interactions between two advanced driver-assistance system features.
Unlike \tool, those system adopted a central design which offered little flexibility.
}

\subsubsection{Software Isolation.}
Partitions have been widely used in safety-critical systems
to separate different applications.
Despite separation kernels, hypervisor and containers are often used to separate execution environment to protect sensitive programs or data.
For example, Liu et al.~\cite{yutaoccs15} proposed a virtualization-based SeCage to protect critical secrets against exploiting memory disclosure vulnerabilities.
Xu et al.~\cite{xu2015condroid} proposed a container-based solution, Condroid, for Android devices to mitigate security risks. 
Xu et al.~\cite{xu-tc} discussed how privileged virtual machine could be affected by hardware errors and how hypervisor can cope with it.
Xiao et al.~\cite{xiao-tc} and Shan et al.~\cite{shan-tc} also used virtualization technology to provide fault isolation by encapsulating application instances.
Thus, here we use unprivileged virtual machines and containers to isolate different ADS components.

%% file: SourceCode/motivation.tex

\section{Motivation}\label{motivation}

ROS~\cite{ros09} is an open-source and~flexible framework for developing robot control systems, which is prevailing in robotics.
Most self-driving systems are built with such a framework,  and the various libraries and packages (contributed by both community users and the industry) built on top of the Linux kernel.
In such a way, the creation of complex and robust robot behaviors across various robotic platforms is simplified.
However, as shown in Fig.~\ref{avsystem}, it also leads to a giant system consisting of tremendous processes interacting with each other to perform complex functions, including sensing, data processing, localization, route planning, motion control, trajectory control, etc., which makes it vulnerable to malware, software failures, or even a kernel panic~\cite{Checkoway2011, koscher2010experimental,jonathan,miller2014survey}.
For example,  malware can be remotely distributed into self-driving cars~\cite{vassallo2018spatially,attackcan} to disrupt system execution or intrude the in-vehicle bus.
Thus, it is urgent and necessary to decouple and isolate each functionality.

In this work, we consider that an adversary may (1) intrude the self-driving system through network connections, such as the V2X communication or the software update over-the-air; (2) implant malware to interfere the system execution, such as consuming available resources, sending fabricated control commands, or even disrupting critical services to cause system failures, which are frequently used in remote attacks~\cite{greenberg2015hackers,vassallo2018spatially}.
Software vulnerabilities may also trigger abnormal execution and lead to failures.
In this work, we assume an ongoing threat caused by either a software failure or attack can lead to anomalies such as malfunctions or abnormal resource usage, which
 can be handled by re-execution or backup-restore.
We also assume the underlying host OS infrastructure (including the host OS kernel and the hypervisor) can be trusted.
No physical attack against ECUs, actuators and sensors is considered.

%% file: SourceCode/decouple_model.tex
\section{Design Principles}\label{designs}

To address the challenges presented in Sec.~\ref{intro}, \tool is designed to be a self-protection framework which can defend the ADS with runtime monitoring and system adaptation.
In \tool, multiple independent guards protect their own target ADS components and work cooperatively, which makes \tool a essentially decentralized system.
The design premises include the following.

First, the effective detection of failures and attacks. 
The protection system should effectively detect failures or attacks in order to maintain the correct execution of the ADS. 
Thus, the protection mechanism should be able to discover anomalies from multiple dimensions, which requires that different detection approaches shall be combined.

Second, the runtime protection of the ADS.
As the ADS consists of various software components, it is essential to keep an ongoing threat (e.g. failure) from propagating and compromising other functionalities/services by isolating the software execution.
Moreover, for all isolated components, as there may be anomalies detected simultaneously, it is critical to correctly locate and mitigate the root cause, instead of blindly shutting down all suspects.

Third, the decentralized system design for better flexibility and reliability. 
The flexibility and scalability require the protection system to be able to upgrade or migrated easily.
For example, as upgrading ADS will change the attack surface by introducing implementation changes,  which requires to upgrade the corresponding protection without pausing other defense functions. 
Besides, it is also important to keep the self-protection systems from single point failures, which remains a challenge in the current central system-level adaptation and reconfiguration design. 
Thus, all above consideration requires the protection scheme to be a decentralized system.

Finally, the overall design 
 of~\tool follows the principles below.
\begin{itemize}[leftmargin=*, topsep=0pt]
\item Decoupling the ADS and isolating its components. 
Under such a principle, the ADS will be divided into a set of components, which will be further isolated by using partitions (Section~\ref{sysmodel}).
\item Local self-protection through service degradation.
Each isolated component is guarded by an independent self-protection loop, which monitors the component and performs necessary operations to repair failures if any (Section~\ref{design}).
\item Global self-protection through cooperation. 
In such a system, 
cascading failures can still be triggered as false input data may paralyze a component following the data flow, which could make the local protection mistakenly executes the protection action against the innocent component.
Thus, it is essential for the protection to locate the root cause component,
which requires a cooperation scheme among all local protection mechanisms.
\end{itemize}
\vspace{-10pt}
\subsection{Component-based Decoupling and Isolation}\label{sysmodel}
In the autonomous driving system, the control performs various functions, and each of them contains several parallel and$/$or sequential processes.
Here, a group of similar processes can be intuitively defined as a component, which decouples the ADS into a set of components. 
Thus, the component model can be presented as a tuple~$(\mathcal{C}, \mathcal{R})$, where $\mathcal{C}$ is a finite set of components performing different functions, and $\mathcal{R} \subset \mathcal{C} \times \mathcal{C}$ is the set of connections among components.

\begin{figure}
	\centering
	\includegraphics[width=0.75\linewidth]{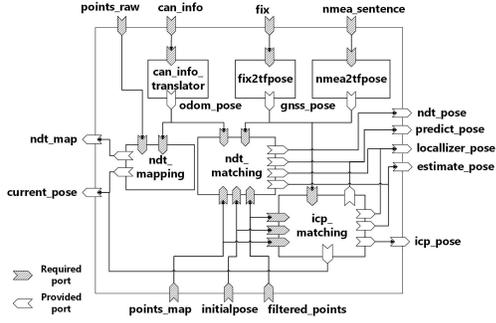}
	\caption{The \emph{Localization} component abstraction of Autoware.}
	\label{comp-dia}
\end{figure}

\begin{definition}\label{def:port}
	A port $p$ is a tuple $(M,\alpha,\mu)$, where $M$ is a finite set of methods in $p$; $\alpha \in \{\text{\emph{provided}, \emph{required}}\}$ is the port type; and $\mu \in \{synchronous, asynchronous\}$ is the communication type.
\end{definition}

\begin{definition}\label{def:component}
	A component $c$ is a tuple $(P_p, P_r, G, W)$ , where~$P_p$ is a finite set of provided ports; $P_r$ is a finite set of required ports; $G$ is a finite set of sub-components; and $W \subset TP \times IP$ is the non-reflexive port relation, where $IP=\cup_{sc\in G} (sc.P_p \cup sc.P_r)$ is the internal ports, and $TP=P_p\cup P_r \cup IP$ is the total ports.
\end{definition}



We have applied both~\eqref{def:port} and~\eqref{def:component} to \emph{Autoware}, and derived the component abstraction for each functionality.
For example, Fig.~\ref{comp-dia} shows the Localization component, which collects data from CAN bus (e.g., can\_info), map (e.g., points\_map) and sensor (e.g., nmea\_sentence) as required ports, and outputs the vehicle position based on Lidar readings, GPS coordinates and odometer feedbacksas provided ports (e.g., current\_pose, predict\_pose).

ADS Components are determined by aggregating similar atomic components. 
A component $c$ is an atomic component if $c.G=\emptyset$. 
A component provides a service to others via a provided port, and requires a service from others via a required port.
Specifically, in a ROS-based system, a functionality may contain several ROS nodes (functional processes). 
Thus, each ROS node can been regarded as an atomic component.
From the above definition, a component is defined by two basic principles.
First, similar functional processes should be put in a component,
if (1) their required ports are similar, such as different localization processes need same data (e.g., gnss\_pose required by both ndt\_matching and ict\_matching nodes as shown in Fig.~\ref{comp-dia}) to produce results with different algorithms; (2) their provided ports are similar, such as providing map data in different forms.
Second, the cardinality of~$\mathcal{R}$ and~$\mathcal{C}$ should be balanced. 
A fine-grained partition of the functionality set could lead to large~$\mathcal{C}$ and $\mathcal{R}$.
Given the computing capability of the control computer and the complexity of system integration, they should be reduced.
Because the larger $\mathcal{C}$ and $\mathcal{R}$ are, the more resources will be consumed to protect all components.
Thus, there are 8 components: Data Loading, Localization, Sensing, Data Fusion, Path Planning, Object Tracking, Motion Planning, and Path Following, which is also derived from a general ADS architecture~\cite{Lin-2018}.






\begin{definition}
	The component model of the ADS is a tuple~$(\mathcal{C}, \mathcal{R})$, where $\mathcal{C}$ is a finite set of components performing different functions, and $\mathcal{R} \subset \mathcal{C} \times \mathcal{C}$ is the set of connections among components.
\end{definition}

\begin{figure}
	\centering
	\includegraphics[width=0.75\linewidth]{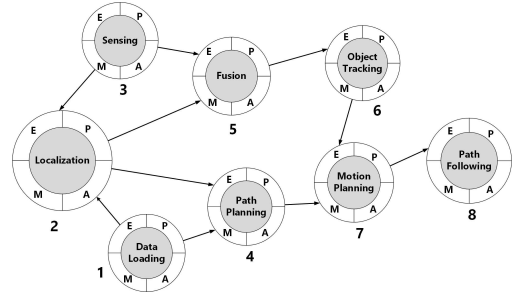}
	\caption{The complete component model of Autoware.}
	\label{dismape}
\end{figure}

	The component model of Autoware is shown in Fig.~\ref{dismape}, where each node denotes a component and arrows compose the set of connections. 
	The component set is $\mathcal{C}$ $=$ $\{$Data Loading, Localization, Sensing, Path planning, Fusion, Object Tracking, Motion Planning, Path Following$\}$. 
	Connection relations describe the data flow among components. 
	For example, Localization component requires data from Data Loading and Sensing components, so $($\emph{Data Loading}, \emph{Localization}$) \in \mathcal{R}$ and $($\emph{Sensing}, \emph{Localization}$) \in \mathcal{R}$.




Given the component model of an ADS $(\mathcal{C},\mathcal{R})$, suppose $\mathcal{C}=\{c_1,c_2,\ldots,c_n\}$, $\mathbb{R}^d$ is the d-dimensional Euclidean space, and $\mathbb{N}$ is the set of all non-negative integers. 
Definition \ref{isoset} defines the general formula of the isolation set of an ADS.

\begin{definition}\label{isoset}
The isolation model of an ADS $(\mathcal{C},\mathcal{R})$ is a tuple $(\mathcal{C},I)$, where $\mathcal{C}$ is the set of components and each of them is put in a partition; and $I$ is a vector-valued function $(r,t,s,b)$ satisfying:
\begin{itemize}[leftmargin=*,topsep=0pt,partopsep=0pt]
\item $r: \mathcal{C} \to \mathbb{R}^d$ specifies the allocation of $d$ kinds of resources~to~the component partitions;
\item $t: \mathcal{C} \to [0,1]$ denotes the isolation level of components, the~higher level of the isolation, the smaller value;
\item $s:\mathcal{C} \to \mathbb{N}$ assigns a mitigation solution to each component, where each number represents a unique policy;
\item $b:\mathcal{C} \to \mathbb{N}$ assigns the maximum allowed number of mitigation operations.
\end{itemize}
\end{definition}



Specifically, $r(c)$ indicates the maximal resources which can be allocated to $c$ (CPU, memory, etc.).
Due to the limited resources, the sum of resource allocated to partitions should not exceed the host available resources.
$t(c)$ represents how $c$ should be isolated and protected.
For simplicity, here  $t(c)$ is set to be either 0 or 1.
$t(c)=0$ means that $c$ is more vulnerable than others, such as it has a larger attack surface or could cause severer failures like a kernel panic, otherwise $t(c)=1$.
$s(c)$ indicates the mitigation policies for $c$.
Currently, only backup and reset operations are considered, so $s(c)$ is set to 0 or 1.
If $s(c)=1$, then the failure of $c$ will be handled by its backups.
Otherwise, the failure will be mitigated by resetting the partition.
$b(c)$ is the allowed number of $s(c)$ operations, which could be the number of backups or allowed resetting operations.
Considering the trade-off between allowed mitigation operations and system performance, for simplicity, only 1 backup or 1 resetting is allowed for each component, 

%% file: SourceCode/adaptation_design.tex
\vspace{-10pt}
\subsection{Self-Protection Mechanism}\label{design}
After the system decoupling,
\tool deploys a local self-protection system for each isolated ADS component. 
Each local protection system operates on its own while cooperating with others to determine the correct adaptation.

\subsubsection{Local Protection mechanism.}
\begin{figure}
	\centering
	\includegraphics[width=0.75\linewidth,height=0.2\textheight]{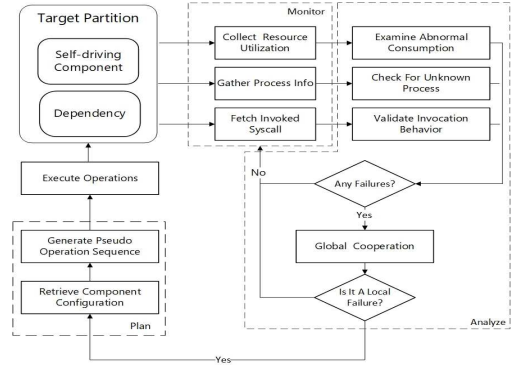}	
	\vspace{-10pt}
	\caption{Local protection process.}
	\label{mape:local}
\end{figure}

Each local protection function, e.g. monitor and planner, is independent and interacts with each other through network communication. 
Such design enables a flexible scheme, allows more functions to be integrated freely in the future, especially new monitoring metrics and corresponding analyzers.
The local protection keeps inspecting the runtime status and behavior of the target, then those gathered metrics are analyzed for any anomaly.
The planner will generate adaptation plan to mitigate the abnormal component according to its adaptation configuration.
Finally, the executor will carry out the planned action to put the target back to normal. 
The whole process is shown in Fig.~\ref{mape:local}.

Currently, the local protection inspects the target component execution with three different approaches, i.e. the resource profiling, process validation and runtime behavior modeling.
Resource exhaustion attacks or potential software bugs may drain computation resources like CPU and memory, which slows down the system and makes service unavailable.
The monitor gathers the resource usage data and sends it to the analyzer periodically.
As it is challenging to set up a usage baseline, the usage inspection is done by examining the current data against the history records from the analyzer side.
A threshold is preset by the system integrator to help the inspection.
Any violation against the threshold will trigger an alert from the analyzer. 

The process validation is designed to identify any rogue process that may harm the system.
The assumption is that the programs running in a component remain the same while various driving data is being processed.
Thus, any newly spawned process could be a potential malware.
As the self-driving system is decoupled within several partitions, the system integrator could identify and determine the allowed user programs and services during test and integration.
Thus, a whitelist could be set for process validation.
The monitor will periodically check for the running process inside the target component partition to gather the process id (PID), name and CR3 control register value.
As PID, CR3 and name should remain unchanged during each execution, the analyzer will validate those values with previous checking result periodically.

To inspect the component execution and identify abnormal behaviors, especially those conducted by hidden processes, a system call (syscall) based modeling technique is adopted.
Syscalls are the prime programmatic way to request privileged services or operations for a user program in Linux systems.
Thus, syscalls tracing is widely used to identify abnormal execution caused by malware, rootkits or software vulnerabilities, etc., especially in host based intrusion detection.
The syscall traces gathered in clean and safe tests could be used to set up the baseline for anomaly detection.
The monitor intercepts all syscall invocations issued by the target component partition in each monitoring window, and extracts required feature data for further check.
The analyzer uses the trained detection model to examine whether the collected invocation data complies with the baseline or not.  

The planner is responsible for generating a mitigation solution.
After the analyzer confirms a rogue partition, the corresponding planner will be notified for further adaptation.
It should first check the component configuration for adaptation options, such as choosing the successor of a failed partition (if multiple backups are available), and then generate a sequence of pseudo actions. 
To restore a component back to normal, two main actions are included.
One is `reset' or `cold restart', which is shutting down (destroying) the component and start it again.
The other is ‘restart’ (or 'warm restart'), which is to re-launch the
component or its backup. 
The planned action will be sent to the executor in pseudo code.
Then the executor will carry out all planned actions by translating those pseudo actions into API invocations to complete the adaptation with the help of the partition manager (e.g. hypervisor or container manager). 
After that, a typical local self-protection loop is completed.

\subsubsection{Global Cooperation mechanism.}\label{unit:interact}
Even though the control system~is~decoupled into a set of partitions, there are still data flows among them. 
Thus, the improper behaviors of a failed component may lead to the failure~of~others, such as producing incorrect results.
This motivates \tool to provide a cooperation mechanism among partitions to identify and perform the adaptation for only the root-cause partition, rather than blindly mitigating all failed ones.
Hereafter, a local protection mechanism and its guarded partition are denoted as a \textit{unit}.
\begin{definition}\label{dfg-df}
	A data-flow graph (DFG) of ADS $(\mathcal{C}, \mathcal{R})$ is a tuple $G=(N,F)$, where
	\begin{itemize}
		\item $N$ is a set of nodes. $\forall n \in N$, $n=(c,~id,~Up,~Down)$ represents a unit, where $c\in \mathcal{C}$ is the component in the unit, $id$ is a unique index integer of the unit, $Up$ is the set of its upstream units, and $Down$ is the set of downstream units;
		\item $F\subset N\times N$ denotes the data flow among units.  
		$\forall~n_1,n_2\in N$, $(n_1,n_2)\in F$ if $(n_1.c,n_2.c)\in \mathcal{R}$. 
	\end{itemize}	 
\end{definition} 

The data flow from unit $A$ to unit $B$ is denoted as $A \to B$,  where $B$ is called a \emph{downstream unit} of $A$ and $A$ is an \emph{upstream unit} of $B$.
All downstream units and upstream units of $A$ form the \emph{Up set} and \emph{Down set} defined in Definition~\ref{dfg-df} respectively.

\begin{ex}
	Fig.~\ref{dismape} shows the control system Autoware with eight units, where the number denotes the unit id.
	Unit \textit{Fusion} requires the data from units \textit{Localization} and \textit{Sensing}, and sends its result to unit \textit{Object Tracking}. 
	Thus, it can be denoted as (\textit{Fusion}, $5$, $\{2, 3\}$, $\{6\}$).
\end{ex}

Upon cooperation, each unit maintains its local knowledge about a detected failure: the set of failed upstream units' $id$s, and the ``sick status'' indicating whether the unit is analyzed as ``failed'' or not.
Such information is exchanged periodically during a cooperative failure handling.

\begin{figure}
	\centering
	\subfigure[State transition of a unit during global cooperation.]{
		\label{unit:state}
		\includegraphics[width=0.7\linewidth]{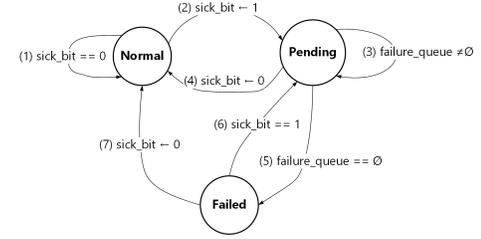}
	}		
	\subfigure[A cooperation example.]{
		\includegraphics[width=0.85\linewidth,height=.2\textheight]{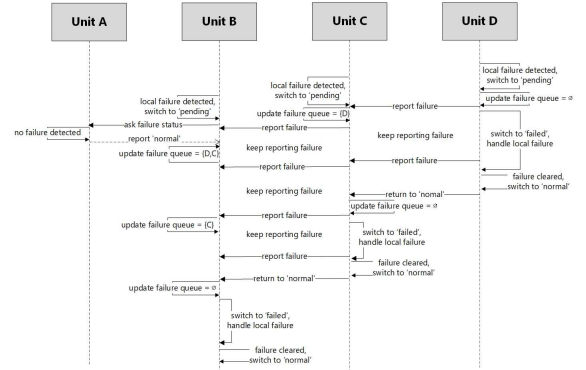}	
		\label{unit:seq}
	}
	\caption{Global cooperation between units.}
\end{figure}

Once the upstream units' failure knowledge is retrieved, a unit determines how it acts, as shown in \figurename~\ref{unit:state}.
Generally, a unit has three states when dealing with possible failures: 1)~\textit{normal}, when the~unit determines that there is no failure; 2)~\textit{pending}, meaning that the unit is waiting for failure assertions from its upstream units; and 3)~\textit{failed}, meaning that the unit is really suffering from a local failure which has to be mitigated. 
Transitions among the states are described in Fig.~\ref{unit:state}. A unit~stays at the \textit{normal} state if no failures are detected (i.e., transition (1)). Once it detects a failure, the unit switches to the \textit{pending} state (i.e., transition (2)) and stays at the \textit{pending} state~if~its $failure\_queue$ (the set of failed upstream units) is not empty (i.e., transition (3)); if $sick\_bit$ (sick status) is 0 (i.e., the former failure is caused by another component), the unit switches back to the \textit{normal}~state (i.e., transition (4)); otherwise, if the $failure\_queue$ is empty, meaning the unit is actually failed, the unit switches to the \textit{failed} state (i.e., transition (5)). After executing the planned adaptation actions, the unit switches to the \textit{normal} state if the failure no longer exists (i.e., transition (7)). Otherwise, it switches back to the \textit{pending} state (i.e., transition (6)).


For example, given a DFG in which $A\rightarrow B \leftarrow C \leftarrow D$, assume that 1) \textit{A and D} have no upstream units, \textit{B} has no downstream units; and 2) \textit{D, C} and \textit{B} suffer from failures while A is healthy.
Then the cooperation between 4 units will be conducted as shown in Fig.~\ref{unit:seq}.
When \textit{D} detects a local failure, it will switch to `pending' state, and reports to its downstream unit \textit{C} for the failure status.
\textit{C} will add \textit{D} into its failure queue and tells its downstream unit \textit{B}.
As \textit{B} is also in `pending' state and is only aware of \textit{C}'s status, it will request the rest of its upstream units, i.e. unit A, for failure status.
After \textit{A} reports no failure detected, \textit{B} will update its failure queue $\rightarrow \{D,C\}$.
When \textit{D} is handling its failure, it will keep report its failure status until the adaptation is done.
Once it is back to `normal' state, \textit{D} will send a failure clear message to notify its downstream units.
Upon \textit{C} receives that notification, it will remove \textit{D} from its failure queue and keeps broadcasting its own updated failure status.
After \textit{B} receives those messages, it will update the failure queue from $\{D,C\} \rightarrow \{C\}$.
As all \textit{C}'s upstream units (unit \textit{D}) are normal, \textit{C} will begin local failure handling and return to `normal'.
After that, \textit{C} will send notification to its downstream unit \textit{B} so that \textit{B} is aware that all units in the failure queue are back to normal now.
Finally, \textit{B} will carry out adaptation to deal with the local failure, then it switches back to ``normal'' state.  

%% file: SourceCode/implementation.tex
\section{\tool System}\label{implementation}
We have implemented the \tool prototype on \emph{Autoware}.
In this section, we will present details on how the system isolation and the self-protection system are achieved.

\vspace{-10pt}
\subsection{System Isolation}\label{sysiso}

\begin{figure}
	\centering
	\includegraphics[width=0.75\linewidth,height=0.12\textheight]{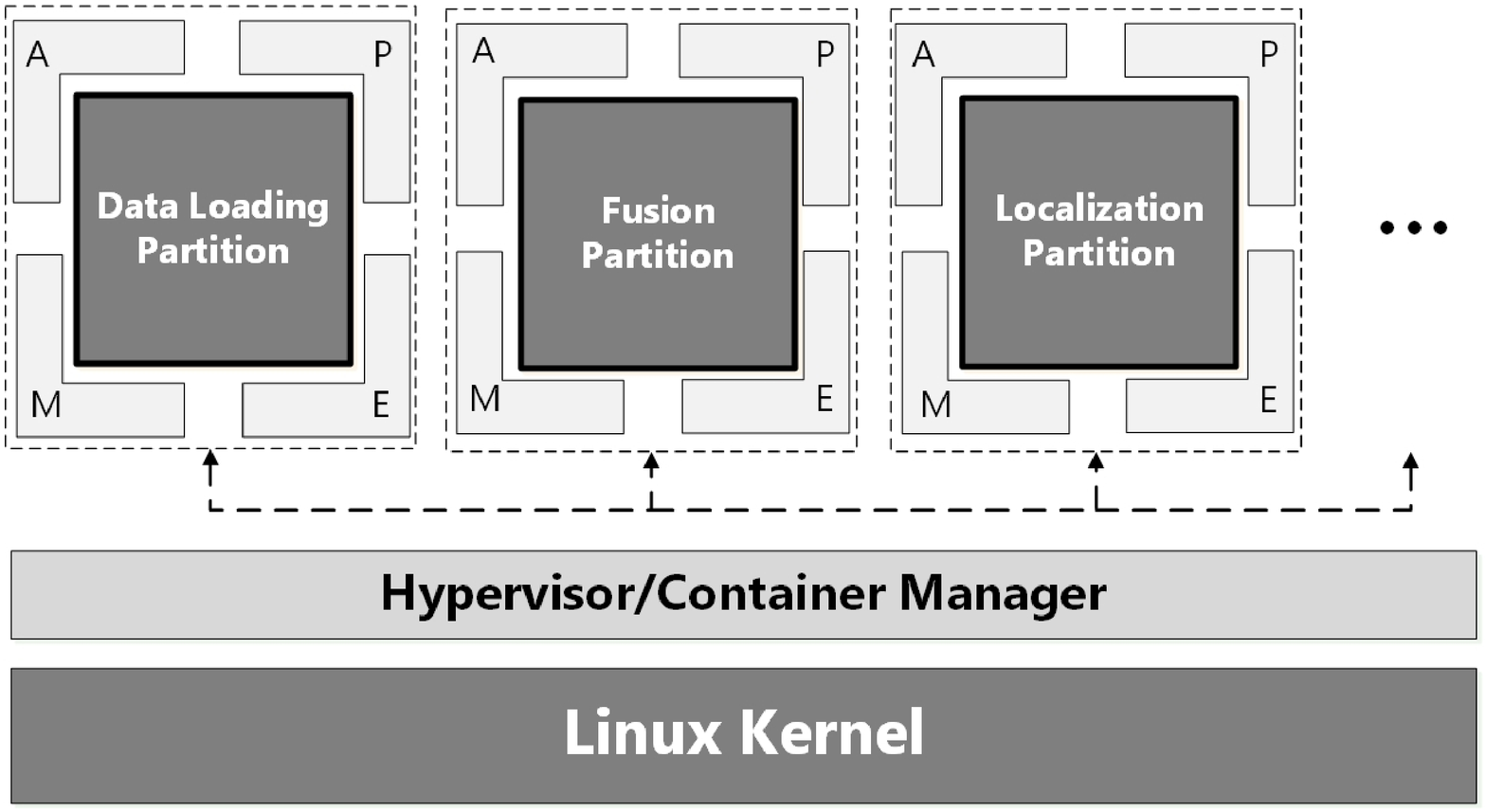}
	\caption{Architecture of \tool.}
	\label{arch}
\end{figure}

As shown in Fig.~\ref{arch}, \tool currently uses a hybrid solution to achieve isolation.
The hardware-assisted virtualization provided by hypervisors like Xen offers better separation, while containerization like LXD provide better performance~\cite{Sharma:2016:CVM:2988336.2988337,vmm_lxc}.
As the safety and security threats may arise from both inside and outside the self-driving functions/software, TrustZone-like solutions are not adopted.
As such solutions offer only one trusted execution environment (secure world), which cannot meet the requirement of separating the self-driving functions from other applications/services 
and the ADS components from each other.
However, a fully verified isolation kernel or microvisor may be leveraged.
But due to the availability and implementation issues, currently we do not find such a capable open system. 
  
Components will be encapsulated based on their properties introduced in Section~\ref{sysmodel}.
If $t(c) = 1$, then the component will be isolated in an unprivileged LXD container, otherwise in an unprivileged hardware virtual machine (HVM).
Given the isolation model introduced in Section~\ref{sysmodel}, the properties of each component in \tool can be determined as shown in Table~\ref{cptable}.
Those accessing devices are put in separated virtual machines because they are more exposed to potential attacks.
Once those components are compromised or failed, the possible crash in either guest kernel or the software will not easily spread and affect others.
Other components are running in unprivileged containers together with necessary ROS libraries and third party applications.
Components including \textit{Fusion, Object Tracking, Path Planning} and \textit{Motion Planning} do not require interaction with external devices as they only rely on other components to feed them input data.
Thus, they are relatively less vulnerable to attacks, and their failures are less likely to cause severe impacts (e.g., kernel panic).
Finally, wireless connections (e.g., wireless interfaces) can also be encapsulated in another dedicated partition with the help of device pass-through, in order to reduce the possible attack surface.

\vspace{-10pt}
\subsection{Local Self-Protection for Each Partition}

\tool is mainly implemented in C++/Python, where a variety of libraries and technology are adopted.

\subsubsection{Monitoring and Analysis.}
\begin{figure}
	\centering
	\includegraphics[width=0.95\linewidth]{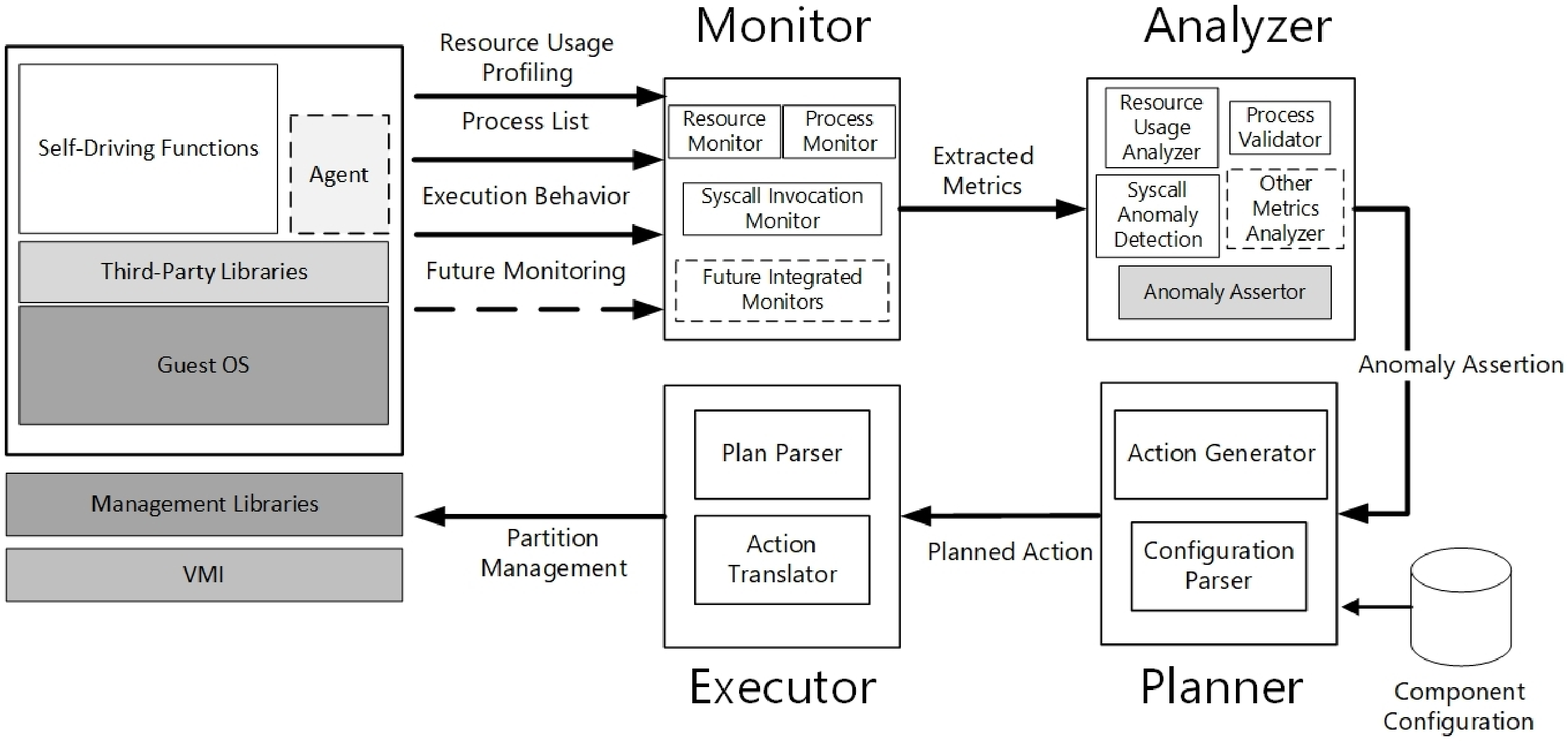}
	\vspace{-5pt}
	\caption{Procedure of component self-protection.}
	\label{mapeproc}
\end{figure}

Monitors inspect target components' behaviors by collecting the resource usage, the running processes and the execution behavior, as shown in Fig.~\ref{mapeproc} (``Monitor'' and ``Analyze'').
Those data are gathered in a clean and safe state to set up the baseline for analysis.
An embedded agent distributed with the base partition image gathers resource usage and transfers it to the monitor.
Compared  with external monitoring, the agent can collect a variety of date accurately and conveniently by using \textit{psutil}.
CPU and memory utilization is examined by preset individual threshold, and is dynamically compared with historical statistics.
The data transmission also serves as heart beats to indicate whether the monitored target is running or not.


\review{2-7}{
The running processes gathered are used for process validation.
To identically label a running process, the PID, the process name and the CR3 register value are retrieved.
With the help of LIBVMI and pre-dumped kernel symbol list, the first step is to locate the memory address of \emph{init\_task}, which could be used to locate the \emph{task list} as Linux stores all processes in such a circular doubly linked list.
To further retrieve the required information such as PID,
according to the offset of PID in \emph{task\_struct} structure (defined in Linux source code include/linux/sched.h), the address of PID could be located and its value can be retrieved as an \emph{int}.
Following the same idea, the name is to locate the \emph{comm} array and get the content.
By walking the \emph{task list}, each process could be examined.
The CR3 is retrieved with syscall invocation and will be explained later.
The process name is checked with the white list as both should be unchanged during any execution.
The PID and CR3 values are recorded when a component is brought up, and both should remain the same in every check during the current execution.
}

\review{2-7}{
The syscall invocation is traced by leveraging breakpoint-based interception~\cite{drakvuf2014}.
Such technique traps syscall entry addresses (located by address translation with the given system symbol map) by injecting \textit{INT3} instruction, so that any invocation in kernel space could be traced, which is efficient against rootkits.
When such a trap is triggered, an event callback will filter all registers to get CR3, record the current process and the invoked syscall.
Given that there are  multiple processes in the target component partition, those data gathered in clean and safe state are used to establish the white list of allowed syscalls, and the frequency feature of invocation pattern is extracted and trained with one-class support vector machine.
During the runtime checking, the analyzer will alert any anomaly invocation against the pre-trained SVM model.
}

CAN bus is used in self-driving cars to provide communication between the control PC and ECUs. 
\tool \ uses \textit{libpcap} and \textit{SocketCAN} to capture CAN frames, then calculates the frequency and build frame event sequence in each sliding window.
As frames of smaller \textit{ids} have higher priority in the bus arbitration, if such frames are transmitted in a high frequency, other transmission is likely to be cut off.
Thus, the frames ids and counts are used to identify any illegal activity.
As the normal CAN frame transmission is highly regular~\cite{ccs2016,miller2014survey}, any sudden frequency or count increase could cause a potential DoS attack.
To double check the system behavior, the captured frames are also cross validated with the control commands issued by \textit{Autoware} to identify any stealthily injected frames.

The analyzer also supports user defined policies, which will be parsed by a flex/yacc-based implementation.
The preprocessor receives monitoring data and parses them into proper data structures, then feeds the processed results into the analysis module.
The analysis module will automatically parse the predefined policy into several individual rules to examine the input data, and score each checking as 0 (not matching) or 1 (matching).
For example, inspired by~\cite{xiaoning2}, to check CAN bus activities in \textit{Path Following}, \tool gathers the event sequence in a sliding time window $h$.
An event $e = \{id, t(id)\}$, where $id$ represents CAN frame id and $t(id)$ is the recorded time stamp.
Thus, an event sequence can be defined as $\rho = [e_1, e_2, ..., e_n, \ldots]$.
Then rules can be defined for this sequence, such as: $p_1$: $(Cx:\langle t(id_1), t(id_{1}) + h\rangle . Cx < \theta_1)$ and $p_2$: $(Cx: \langle t(id_1), t(id_{1}) + h\rangle . Cy: \langle t(id_2), t(id_{2}) + h). (Cx - Cy) < \theta_2)$.
where $Cx$ is the number of $id_1$ in the time window $h$ starting from $t(id_1)$, and $Cy$ is the count of $id_2$.
$p_1$ checks single frame count, while $p_2$ checks any related two sequences.
Multiple rules could be integrated like $p_1  \&\&\  p_2 \ \&\& ... \&\&\ p_m$, which is used to perform complex checking.
After that, scores are used to determine if there is something suspicious.
A score voter compares the results from a successive number of sliding windows to see if the majority fail or not.

\subsubsection{Plan and Execution.}
Once the planner receives a failure report from the analyzer, it will check the corresponding adaptation configuration to plan possible mitigation as shown in Fig.~\ref{mapeproc}.
According to each component's properties, planners take different actions.

\begin{itemize}[leftmargin=*,topsep=0pt,partopsep=0pt]
\item For a component with no backups ($b(c)=0$), planners will reset it.
\item For those whose $b(c)=1$, planners will replace them with their backups.
\item Otherwise, planners call for a system reboot to put everything back to normal.
\end{itemize}


\tool \ uses XML-based configuration files to store a component's adaptation settings.
Each configuration file contains a ``node'' (the main component) and its ``sub-nodes'' (backups).
Each $node$ is sorted by its $id$, and each main component's $id$ is 0.
If multiple backups are allowed, then the sub-node with $id=n+1$ will be the successor, where $n$ is the id of the current running one.
Upon a restoring decision is made, planners will generate the pseudo action sequence for the executor, e.g. planners will first boot a backup, then launch necessary services, and finally shut down the original partition.
Each plan message includes the following fields.
\begin{itemize}[leftmargin=*,topsep=0pt,partopsep=0pt]
	\item $header$, containing sequence id, time stamp, and type called `plan' which indicates it is a plan message;
	\item $opcode$, indicating the operation type, such as start, stop, launch or pause;
	\item $successor~name$, telling the backup partition's name, which is retrieved from the configuration file;
	\item $launch\_file\_path$, telling how to launch the backup service if needed, retrieved from the configuration file;
	\item $payload$, other arguments like the IP address and partition type, retrieved from the configuration file.
\end{itemize}


Those plan actions will be sent serially, and executors could just ``translate'' those sequences into commands or API invocations, then carry them out.
For partition operations including \textit{start, stop, reboot} and \textit{pause}, the corresponding hypervisor or container manager APIs such as \textit{libxenlight} or \textit{pylxd} will be invoked.
If a planner sends \textit{launch} messages to start a service in a partition, then $roslaunch$ python APIs are to be used, the executor refers to the partition name (as host name) to specify where the service process should be spawned by invoking $roslaunch$.


\vspace{-10pt}
\subsection{Global Self-Protection via Unit Cooperation}\label{gsp}

\begin{algorithm}[t]
	\scriptsize
	\algtext*{EndWhile}
	\algtext*{EndIf}
	\caption{Cooperation procedure of unit $u$.}
	\label{alg:unit}
	\algnewcommand{\LeftComment}[1]{\Statex \(\triangleright\) #1}
	\begin{algorithmic}[1]
		\Require
		\Statex $failure\_queue \leftarrow \emptyset$ \Comment{(local) the sequence of failure units.}
		\Statex $sick\_bit \leftarrow 0$ \Comment{the node status, 0: normal and 1: failure.}
		
		\Ensure
		\Statex $failure\_queue == \emptyset$
		\Statex $sick\_bit == 0$
		\Statex 
		
		\Loop  \label{p21}
		\State analyze\_monitoring\_data \label{MA}
		\If {failure detected}\Comment{switch to pending state}
		\If {$sick\_bit == 0$} \label{ntop}
		\State $sick\_bit \leftarrow 1$
		\State ask upstream nodes for failure assertion  	\Comment {in case of beacon delay}			
		\EndIf \label{ntop1}
		\State update $failure\_queue$ by subscribing upstream units'	failure beacons\label{update}		
		\If{$failure\_queue \neq \emptyset$} \label{sp}
		\State publish failure beacon to its downstream units \label{send1}
		\Else \Comment{switch to failure state} \label{ftof}
		\State perform local self-protection \label{ftof1}
		\EndIf
		\Else \label{fton}\Comment{ switch to normal state}
		\If {$sick\_bit == 1$}\label{clear1}
		\State $sick\_bit \leftarrow 0$
		\State $failure\_queue \leftarrow \emptyset$
		\State publish a clear beacon to its downstream units\label{clear2}  		
		\EndIf
		\EndIf\label{fton1}
		\EndLoop \label{p22}
		
	\end{algorithmic}
\end{algorithm}
Unlike a centralized adaptation control which requires a global decision maker, each unit in \tool \ acts independently.
A unit cooperates with others by collecting analysis reports from the upstream and sending its local assertion to its downstream units.
The data structure of each unit includes: 1) a string indicating its \textit{name}, 2) a unique integer as its \textit{id}, 3) queues containing its \textit{upstream} units' $id$s and \textit{downstream} units' $id$s respectively.
Hence, messages traversing among units contain the following fields:
\begin{itemize}[leftmargin=*,topsep=0pt,partopsep=0pt]
	\item $header$, containing message sequence number, time stamp, and type called `unit' indicating it is for cooperation;
	\item $source\_id$, the $id$ of the sender unit;
	\item $sick\_bit$, a boolean value denoting the status of the sender unit;
	\item $failure\_queue$, an array of failed units' $id$s.
\end{itemize}

Each unit publishes its $failure\_beacon_{id}$ in the above format upon any failures detected, and cooperates according to Algorithms~\ref{alg:unit}.
When a failure is detected, the unit enters \textit{pending} state.
It updates its $sick\_bit$ and actively inquires that upstream units (Lines \ref{ntop} $-$ \ref{ntop1}) (if an upstream unit's beacon is delayed or the upstream unit functions well).
Then it updates its $failure\_queue$ according to acquired failure beacons (Line \ref{update}).
If an upstream unit works normally, its $id$ should be removed, otherwise its $id$ should be inserted into  $failure\_queue$.
If $failure\_queue$ is not empty, the unit stays at the \textit{pending} state and keeps sending its $failure\_beacon$ (Lines \ref{sp} and \ref{send1}).
Otherwise, if the failure still exists, the unit confirms it and switches to the \textit{failure} state, then executes the local mitigation (Line \ref{ftof1}).
Once the failure status has been cleared, the unit goes to the \textit{normal} state, it also updates the $sick\_bit$ and $failure\_queue$ to inform the downstream units (Lines \ref{fton} $-$ \ref{fton1}).

%% file: SourceCode/evaluation_tii.tex
\section{Evaluations}\label{evaluations}
In the following test, real driving data~\footnote{\review{5-6}{\url{https://www.autoware.ai/}}} recorded by Tier IV, Inc. (the maintainer of \textit{Autoware}) was used as system inputs, which was recorded during road tests in Moriyama, Japan.
Those data include 3D maps, LIDAR/CAMERA/GPS signals, which were fed into Data Loading, Sensing and Localization components.

\review{2-1}{
Following our design and implementation, we isolated and deployed Autoware  (version 1.5.1) in 9 partitions by using Xen 4.10 and LXD 2.21 as discussed in Section~\ref{sysiso}.
Partition properties are shown in Table~\ref{cptable}.
\tool ran in Xen Domain-0 as it was assumed to be trusted.
Our evaluation platform was a Dell workstation with an Intel Xeon E5-1650 v3 CPU (12 logic cores), 32 GB memory and 1 TB hard drive.
The host operating system was Ubuntu 16.04.3 amd64, with Linux kernel version 4.4.108.
}

\begin{table}
	\scriptsize
	\centering
	\caption{Examples of Component Properties}
	\vspace{-5pt}
	\begin{tabular}{c|c|c|c} \hline
		Partition 				& $t(c)$ 	& $s(c)$ & $r(c)$ \\\hline
		Sensing 				& 0 		& 0 	 & 2 cores, 2GB Memory \\\hline
		Data Loading			& 0 		& 1 	 & 2 cores, 2GB Memory\\\hline
		Localization 			& 0 		& 1 	 & 2 cores, 2GB Memory \\\hline
		Fusion 					& 1 		& 1 	 & 4 cores (shared), 2GB Memory \\\hline
		Object Tracking 		& 1 		& 0  	 & 4 cores (shared), 2GB Memory\\\hline
		Path Planning 			& 1 		& 1 	 & 4 cores (shared), 2GB Memory \\\hline
		Motion Planning 		& 1 		& 1 	 & 4 cores (shared), 2GB Memory \\\hline
		Path Following			& 0 		& 0 	 & 2 cores, 2GB Memory \\\hline
		ROS core node 			& 1 		& - 	 & 4 cores (shared), 2GB Memory \\\hline
	\end{tabular}
	\label{cptable}
\end{table}

\vspace{-10pt}
\subsection{Evaluation on Local Self-Protection}\label{localeval}
We set up two test scenarios to evaluate how \tool could effectively defend the ADS with local self-protection mechanisms.
In the first scenario, an adversary breached and launched an attack in the \textit{Localization} partition to degrade the performance by exhausting the available memory, which delayed the calculation and transmission of corresponding positioning message $/gnss\_pose$.
In the second test, a software failure was injected to crash the \textit{Fusion} component, which cut down the computation and transmission of fused point images $/points\_image$.
Each test lasted for 200 seconds, where either the attack or failure was injected at the 80th second.
In both cases, \tool \ was supposed to detect the abnormal situation and planned a system reconfiguration by starting the backup partition and stopping the compromised service.



\begin{figure}
	\centering
	\subfigure[Localization Failure protection.]{
		\label{attacktest}
		\includegraphics[height=0.365\linewidth]{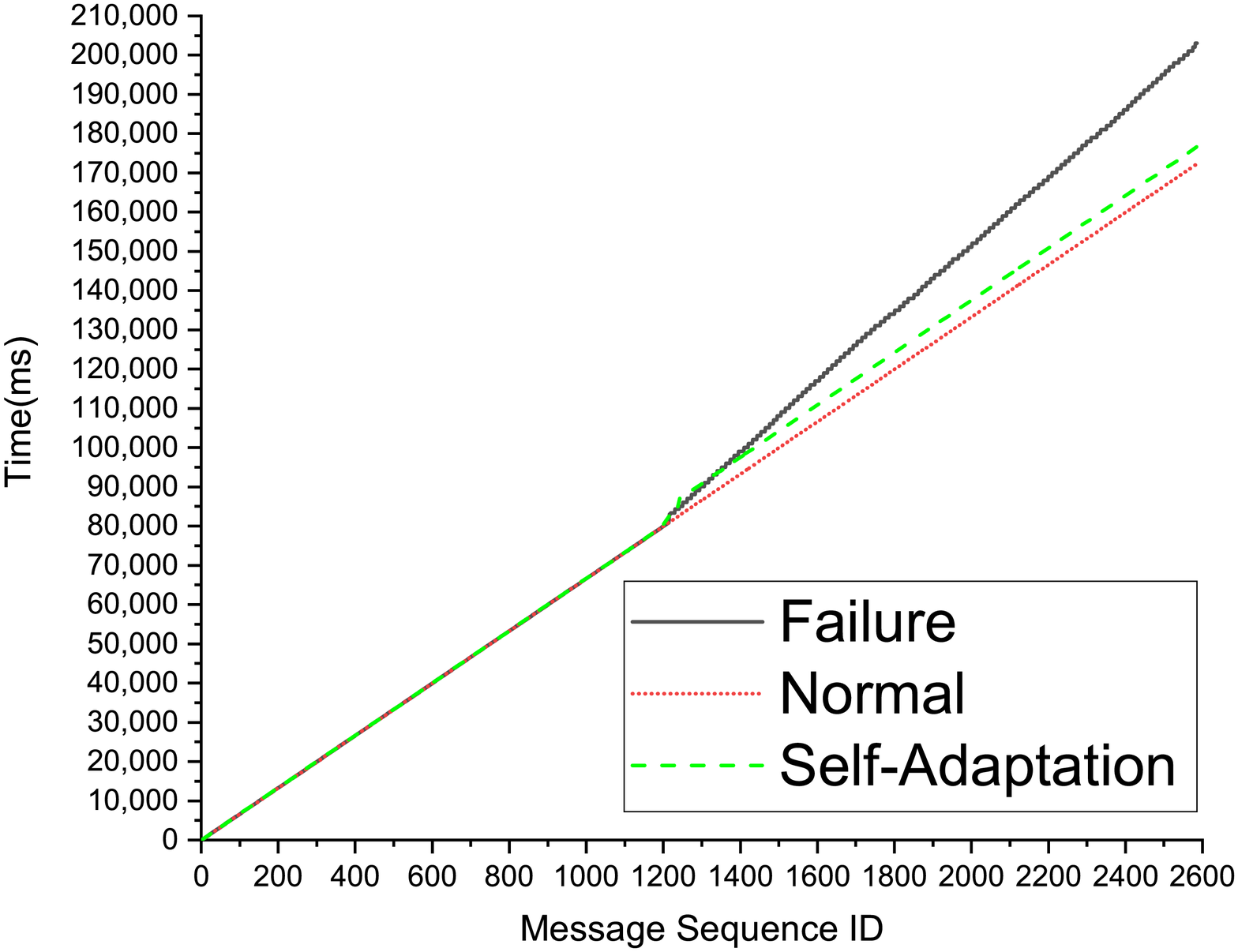}
	}
	\hspace{-1em}
	\subfigure[Fusion crash mitigation.]{
		\label{fusioncrash}
		\includegraphics[height=0.365\linewidth]{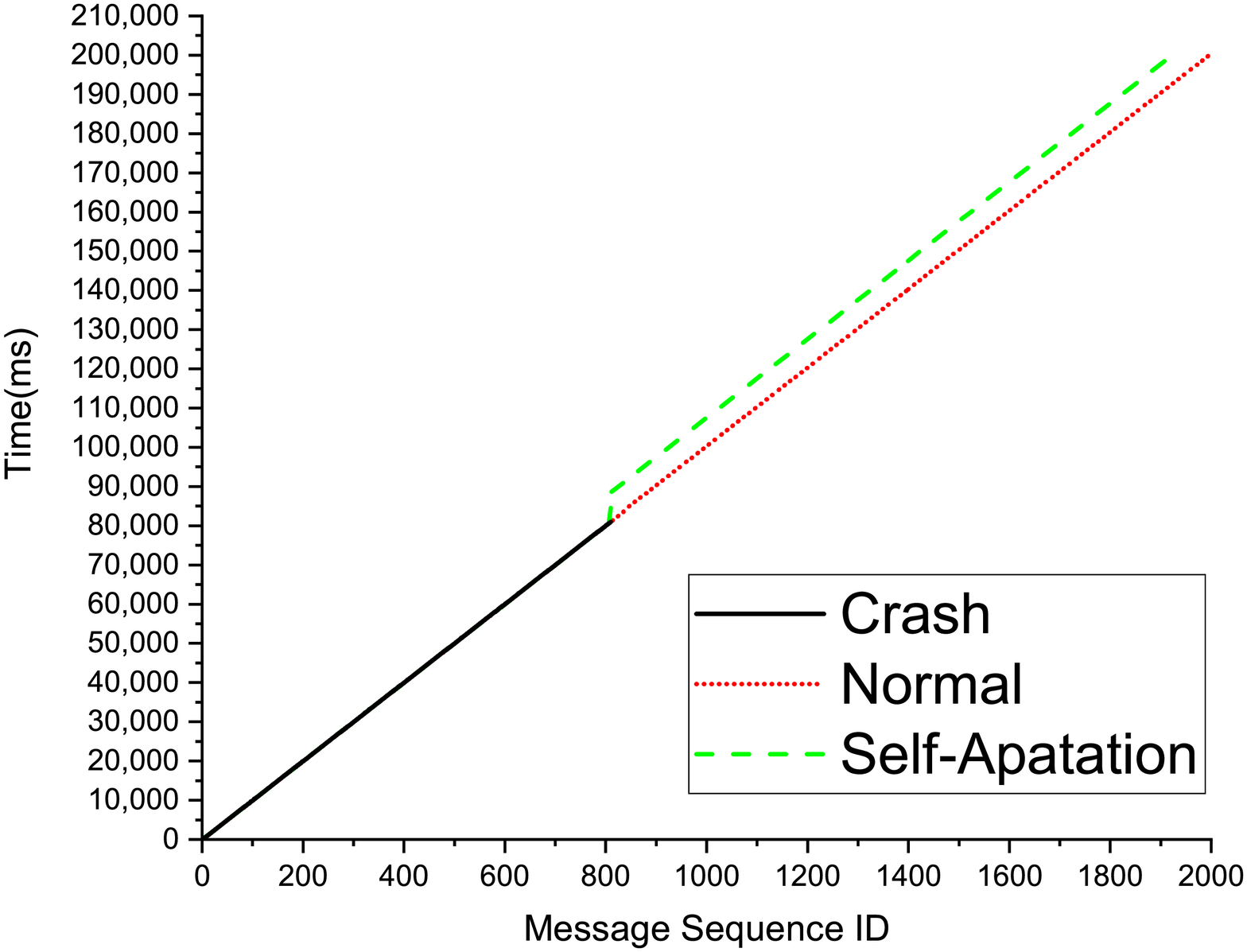}
	}
	\vspace{-5pt}
	\caption{Local protection against failures.}
	\label{localtest}
\end{figure}

Fig.~\ref{localtest} showed the test result, where the $x$-axis denoted
the sequence of messages in time order, and the $y$-axis denoted the sending time stamp.
\review{4-8}{
In Fig.~\ref{attacktest}, as the $Noraml$ curve showed, position messages were transmitted at about every 75ms averagely (the gradient of line).
After the attack occurred at the 80th~second, without \tool, the message transmission was delayed as the solid curve (Failure) showed.
Instead, \tool\ mitigated the abnormal and restored the normal transmission, as the dash curve (Self-Adaptive) was parallel to the dot line (Normal) after the adaptation.
In Fig.~\ref{fusioncrash}, as the component crashed after 80 seconds (the black line ended at the 80th second), \tool \ detected the failure and brought up a backup.
As \textit{Data Fusion} processed large amount of image data, the service restoration took longer (as the dash line showed).
}

As recorded in the test, the cold partition restoration averagely took about 4.64s, recorded from the execution operation began to the first message was received from the backup, including the creation of a partition, the bootstrapping of the operating system and services, and the initialization and computation of the new processes in backup which varied according to their workloads.
It further motivated us to investigate the time overhead and possible future improvements.



To investigate the time overhead, we analyzed the booting of a cold backup in \tool.
First, we analyzed the time of starting a partition. 
It took 466.82ms to boot a container partition from invoking the executor's \textit{start\_partition} function to the first partition status check returning ``Running'' in 100 tests.

Second, we evaluated the time required to start a new ROS node after the partition was ready.
The result showed the $nmea2tfpose$ node of $Localization$ component was launched and initialized in 639.19ms averagely, which suggested that using a hot backup could much further reduce the restoration.

\begin{figure*}
	\centering
	\subfigure[c][DFG of Autoware.]{
		\label{topo}
		\includegraphics[width=0.17\linewidth]{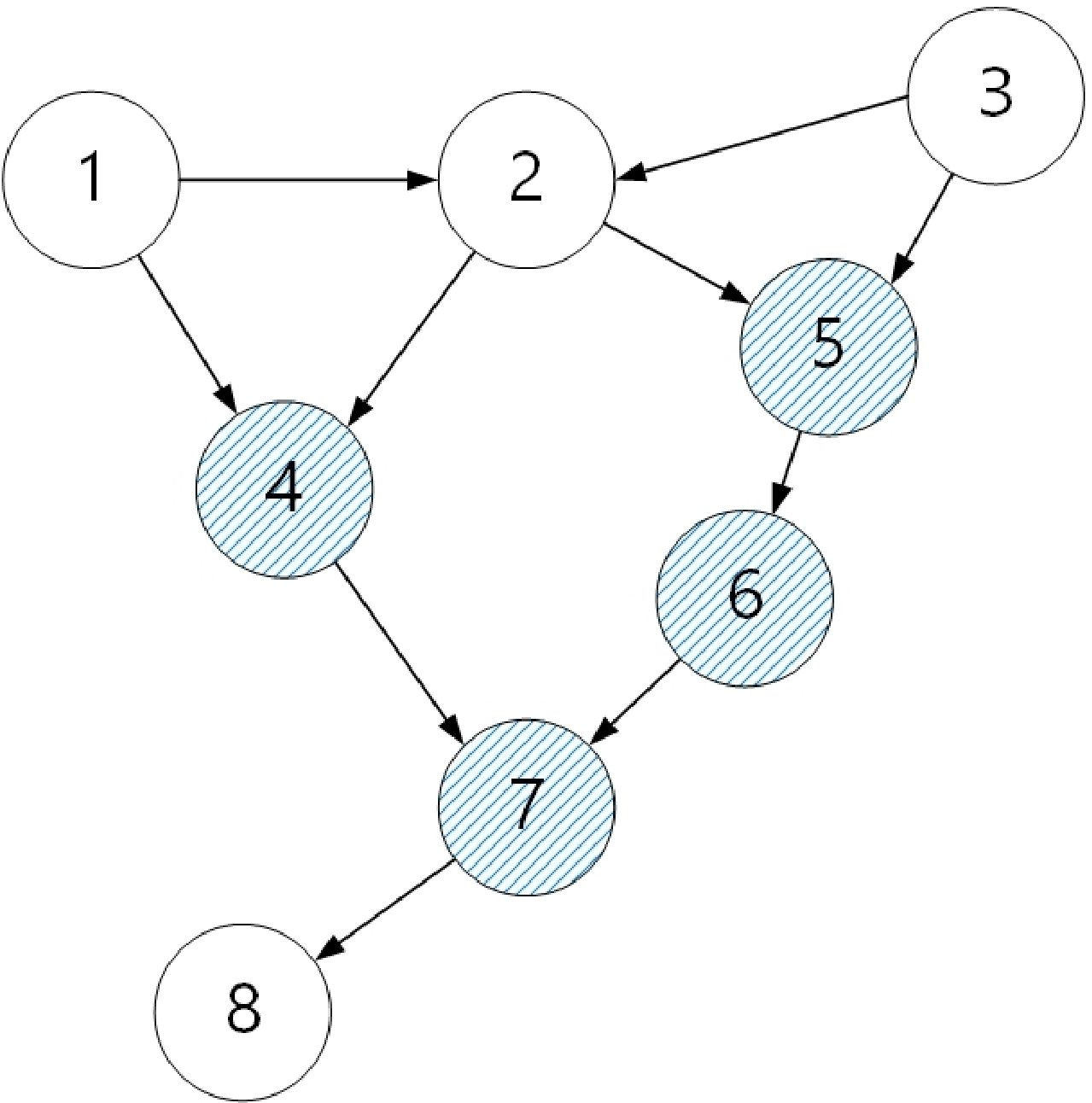}}
	\hspace{0.7cm}
	\subfigure[Failure caused by unit 5.]{
		\label{dis5}
		\includegraphics[height=0.23\linewidth]{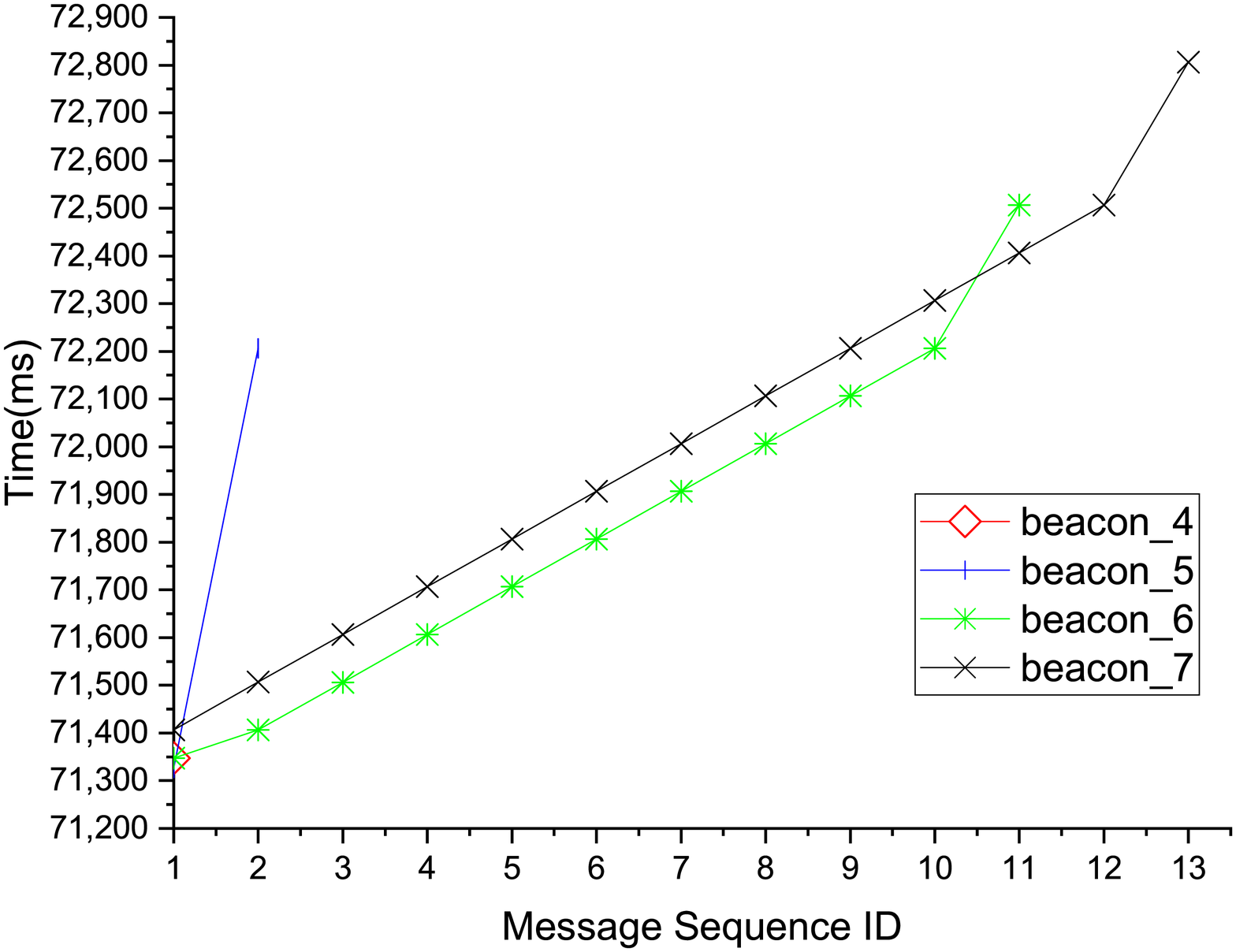}}
	\hspace{0.3cm}
	\subfigure[Failure caused by units 4, 5.]{
		\label{dis45}
		\includegraphics[height=0.23\linewidth]{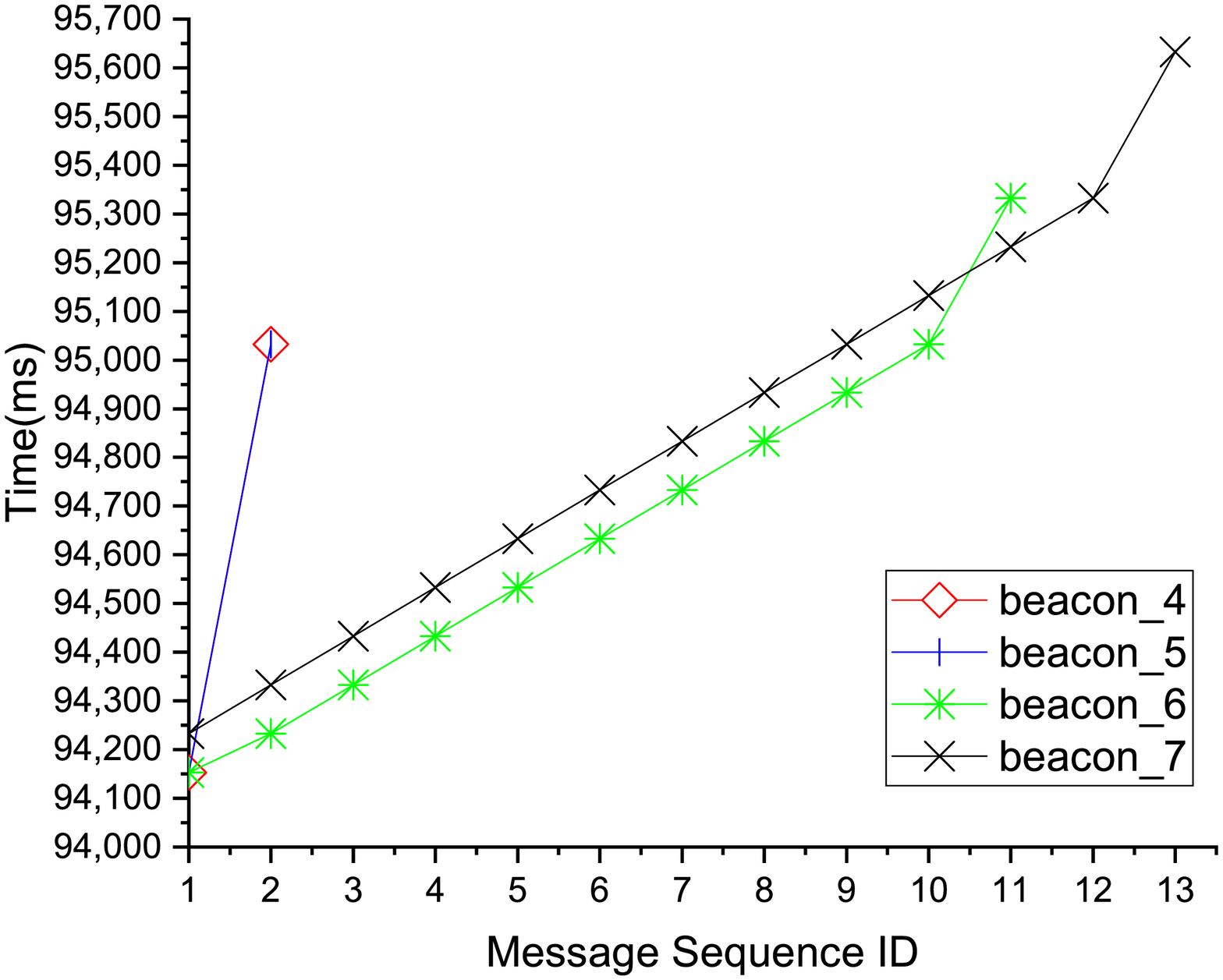}}
	\subfigure[Failure caused by units 5, 6.]{
		\label{dis56}
		\includegraphics[height=0.23\linewidth]{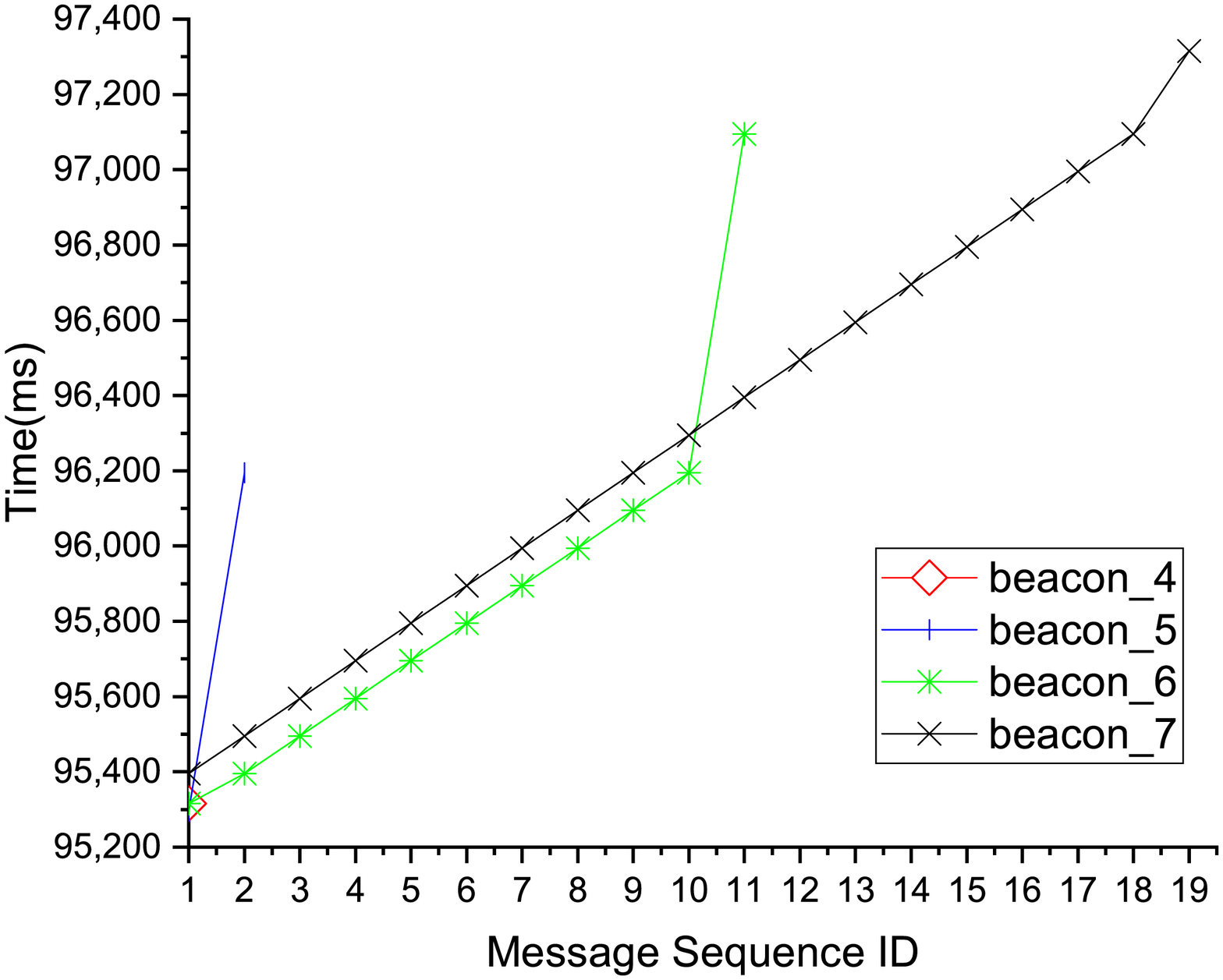}}
	\subfigure[Failure caused by units 4, 5, 6.]{
		\label{dis456}
		\includegraphics[height=0.23\linewidth]{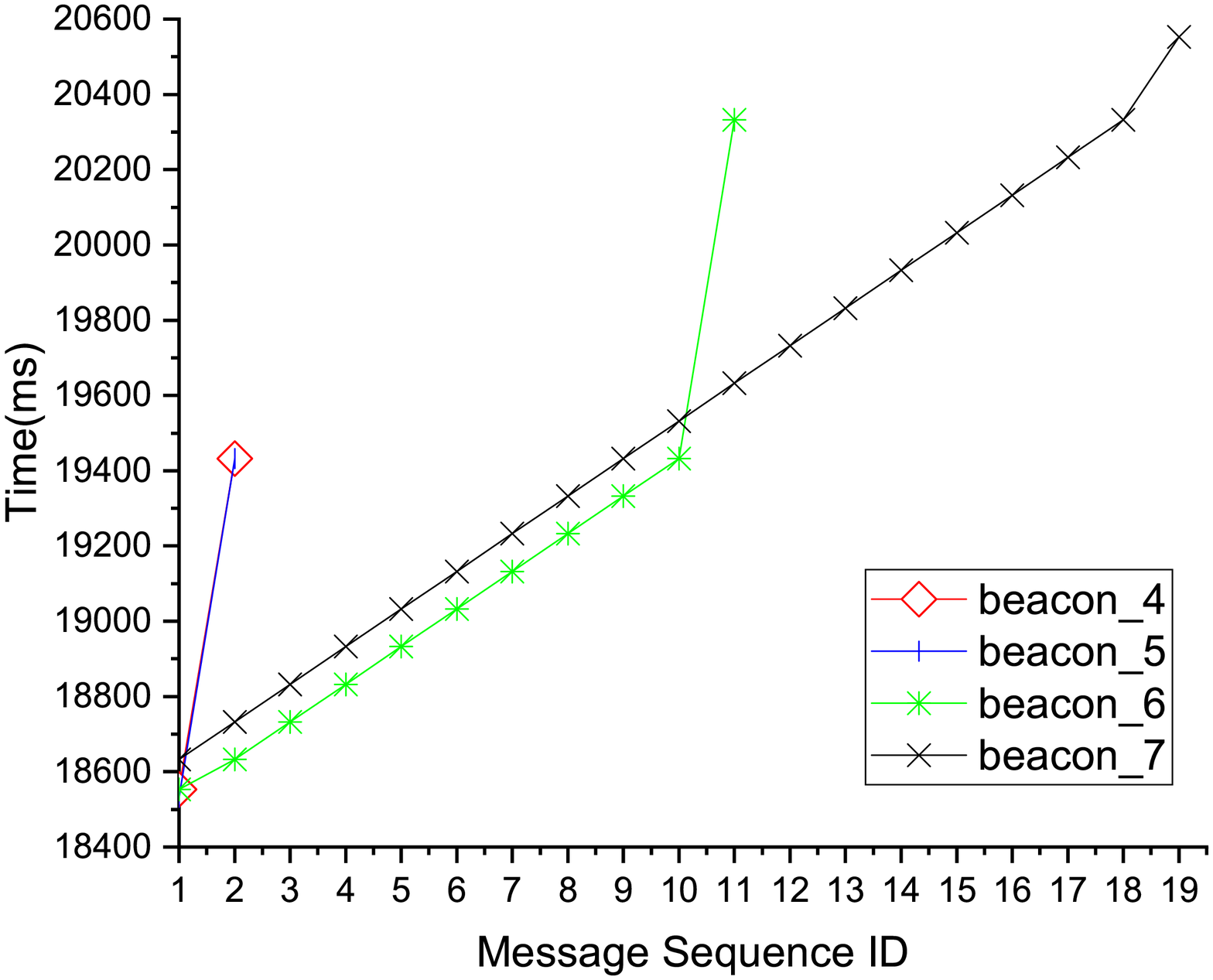}}
	\subfigure[Failure caused by units 4 - 7.]{
		\label{dis4567}
		\includegraphics[height=0.23\linewidth]{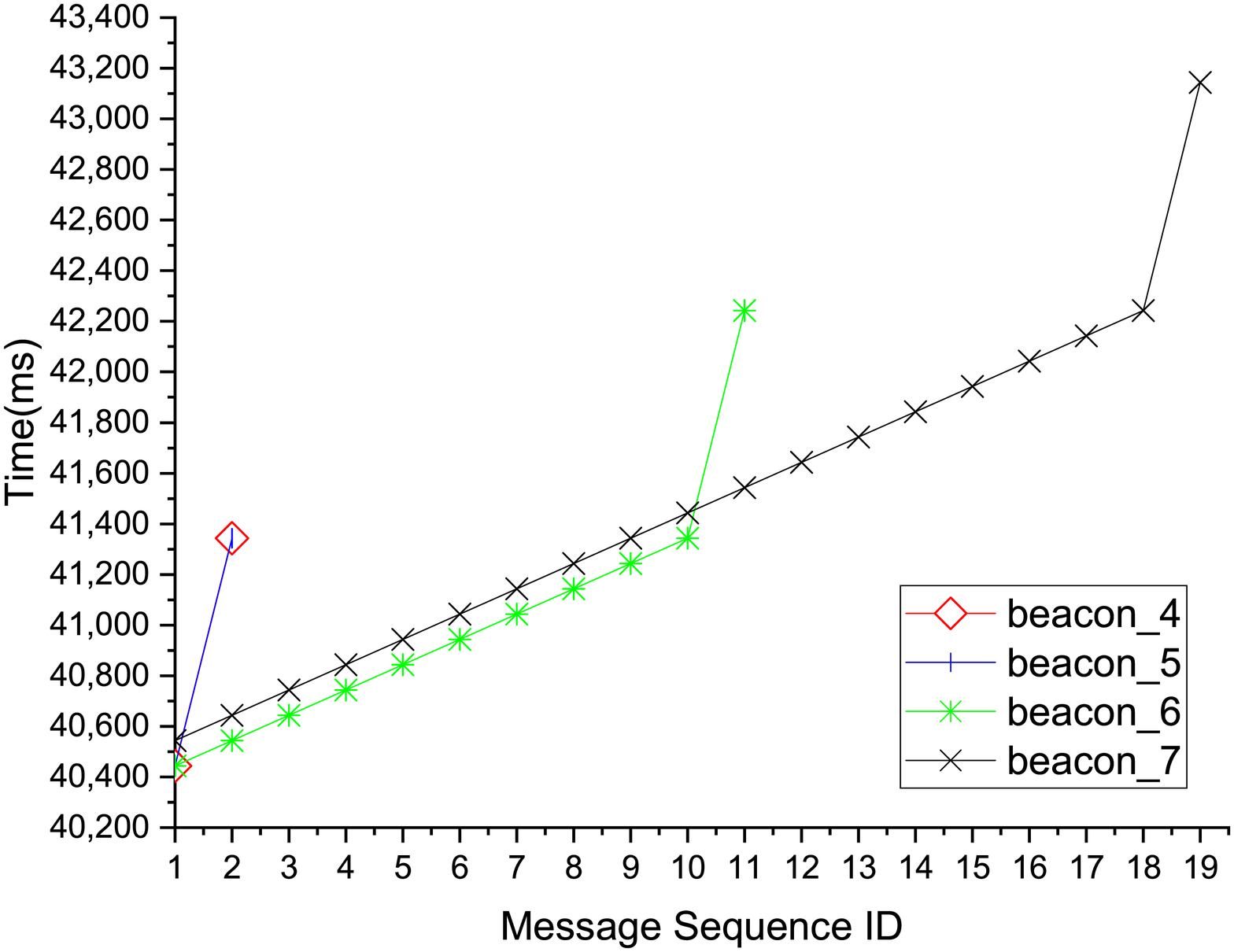}}
	\vspace{-5pt}
	\caption{Global cooperation against multiple failures.}
	\label{discoop}
	\vspace{-10pt}
\end{figure*}

It also indicated that 3534ms (i.e., 4640 $-$ 466.82 $-$ 639.19) was spent in handling other operations. 
Except for booting other OS services (e.g., networking and ssh), computing the workload and publishing the first message (by the node), we found the most of time was spent on handling the race condition of concurrent container operations (by the \textit{lxd} manager).
As the executor of \tool~replaced a component by requesting to start the backup first and then to stop the failed partition, \textit{lxd} manager handled both operations and triggered a race condition on accessing container storage pool and invoking \textit{system calls} in a short period.
It can be optimized by slightly delaying the stop partition request, so the \textit{lxd} manager could handle operations serially, which greatly reduced the time gap from thousands to hundreds of milliseconds as observed in the evaluation.
Meanwhile, the old services will be deactivated replaced by the new ones, so we did not have to worry about data conflicts.


\vspace{-10pt}
\subsection{Evaluation on Global Cooperation}\label{globaleval}
The global cooperation is the key to the decentralized self-protection mechanism in \tool.
Thus, we simulated multiple failures to evaluate the cooperation mechanism in this test. 
To achieve that, we set up
five scenarios where failures were simultaneously occurred in multiple units, i.e., $\left\lbrace5\right\rbrace$, $\left\lbrace4,5\right\rbrace$, $\left\lbrace5,6\right\rbrace$, $\left\lbrace4,5,6\right\rbrace$ and $\left\lbrace4,5,6,7\right\rbrace$,
due to dependency shown in Fig.~\ref{topo}, units $\left\lbrace5,6,7\right\rbrace$, $\left\lbrace4,5,6,7\right\rbrace$, $\left\lbrace5,6,7\right\rbrace$, $\left\lbrace4,5,6,7\right\rbrace$ and $\left\lbrace4,5,6,7\right\rbrace$ were affected respectively.
Based on results and analysis in Sec.~\ref{localeval}, we assumed 650ms was required for the hot backup-restore.
In each simulation scenario, the cooperation beacon message was transmitted every 100ms.
The sending time stamps of all beacon messages were recorded and used to analyze the cooperation performance.

The result showed that the root-cause partitions were correctly identified and handled in all tests.
Besides, Fig.~\ref{dis5}--\ref{dis4567} showed the time overhead of state transitions, where the $x$-axis denoted the sequence of sent beacon messages, and the $y$-axis denoted their sending time stamps.
Specifically, in Fig.~\ref{dis5}, after unit 5 detected its failure,  it spent 879.56ms ($ts(5.2) -ts(5.1)$, where $ts(i.j)$ indicated the sending time stamp of $j$-th message from unit $i$) to complete the local adaptation or send the \textit{clear} beacon message (i.e., setting 0 to $sick\_bit$).

Once unit 6 received unit 5's clear beacon, it spent 299.92ms $(ts(6.11)- \max(ts(5.2), ts(6.10)))$ to prepare and send its own clear beacon. Similarly, unit 7 spent 299.96ms $(ts(7.13)-\max(ts(7.12), ts(6.11), ts(4.1)))$  to complete the above procedure.
Note that unit 4 also sent a beacon message to unit 7 on unit 7's demand.


Similarly, in Fig.~\ref{dis45}, units 4 and 5 took 879.55ms and 879.57ms to adapt to the failures respectively. 
After that, it took 299.96ms for unit 6 and 300.02ms for unit 7 to clear their failure status.
Note that the curves of units 4 and 5 were overlapped as failures simultaneously occurred in both units.


From the above results, it averagely cost 893.0ms for a failed unit to collect beacon messages, complete local restoration and send the clear beacon, while an affected unit took 296.1ms to receive beacon messages and clear its failure status.
Such time overhead could also be reduced by optimizing the beacon transmission speed.
Such time analysis also fit the rest test cases, which indicated that the cooperation mechanism worked stably.
For example, when handling failures caused by one single component, e.g., in Fig.~\ref{dis5}, it took 1479.6ms for all affected units to complete local self-protection actions, i.e., 893.0 (unit\ 5) + 296.1 (unit 6) + 296.1 (unit 7) $\approx$ 1479.6.
Then, the time increased to 2020.0ms ($\approx$ 893.0 (units 4 and 5) + 893.0 (unit 6) + 296.1 (unit 7)) to handle failures caused by $\left\lbrace 4,5,6 \right\rbrace$ and 2699.3ms ($\approx$ 893.0 (units 4 and 5) + 893.0 (unit 6) + 893.0 (unit 7)) for $\left\lbrace 4,5,6,7 \right\rbrace$ (in the above analysis, unit 4 and 5 had no dependency).



\vspace{-10pt}
\subsection{Evaluation on Performance Overhead}\label{penaltyeval}
\review{4-2}{
Most programs in Autoware operate as 1) listening for incoming data,
 2) executing programmed computation logic with those data to output results, 3) sending the result to other interested program. 
Thus, Autoware relies heavily on the network communication.
Further evaluation on the performance overhead imposed on the message transmission latency was conducted.
In this test, the latency data was gathered and analyzed, as it could tell how the program's execution and communication would be delayed, which implied how much overhead could be introduced to the overall system.  
}

First, we ran $Autoware$ in a bare-metal platform and analyzed message latency without \tool to establish the baseline, and then we enabled \tool and reran the experiment.
Each test lasted for 100 seconds and was conducted for 3 times.
The message transmission latency was calculated based on the recorded sending and receiving time stamps.
We focused on three messages with different workloads: localization message (in \textit{Localization}, 136B/each), object image message (in \textit{Object Tracking}, 1050B/each), and point image message (in \textit{Fusion}, 7MB/each).

\begin{figure*}[t]
	\centering
	\subfigure[Localization message latency.]{
		\label{localcomm}
		\includegraphics[width=0.28\linewidth]{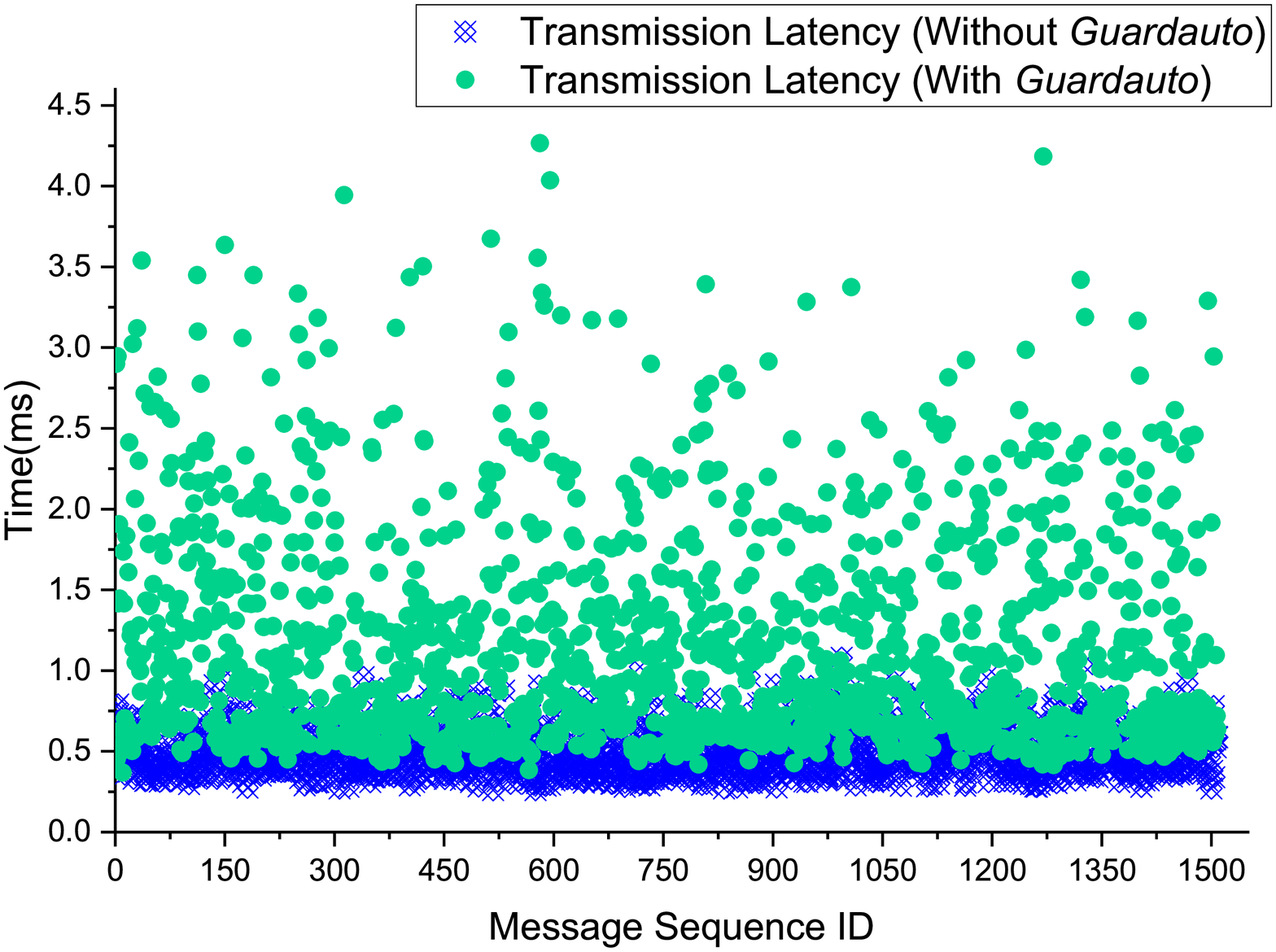}}
	\hspace{0.12cm}
	\subfigure[Object image message latency.]{
		\label{objcomm}
		\includegraphics[width=0.28\linewidth]{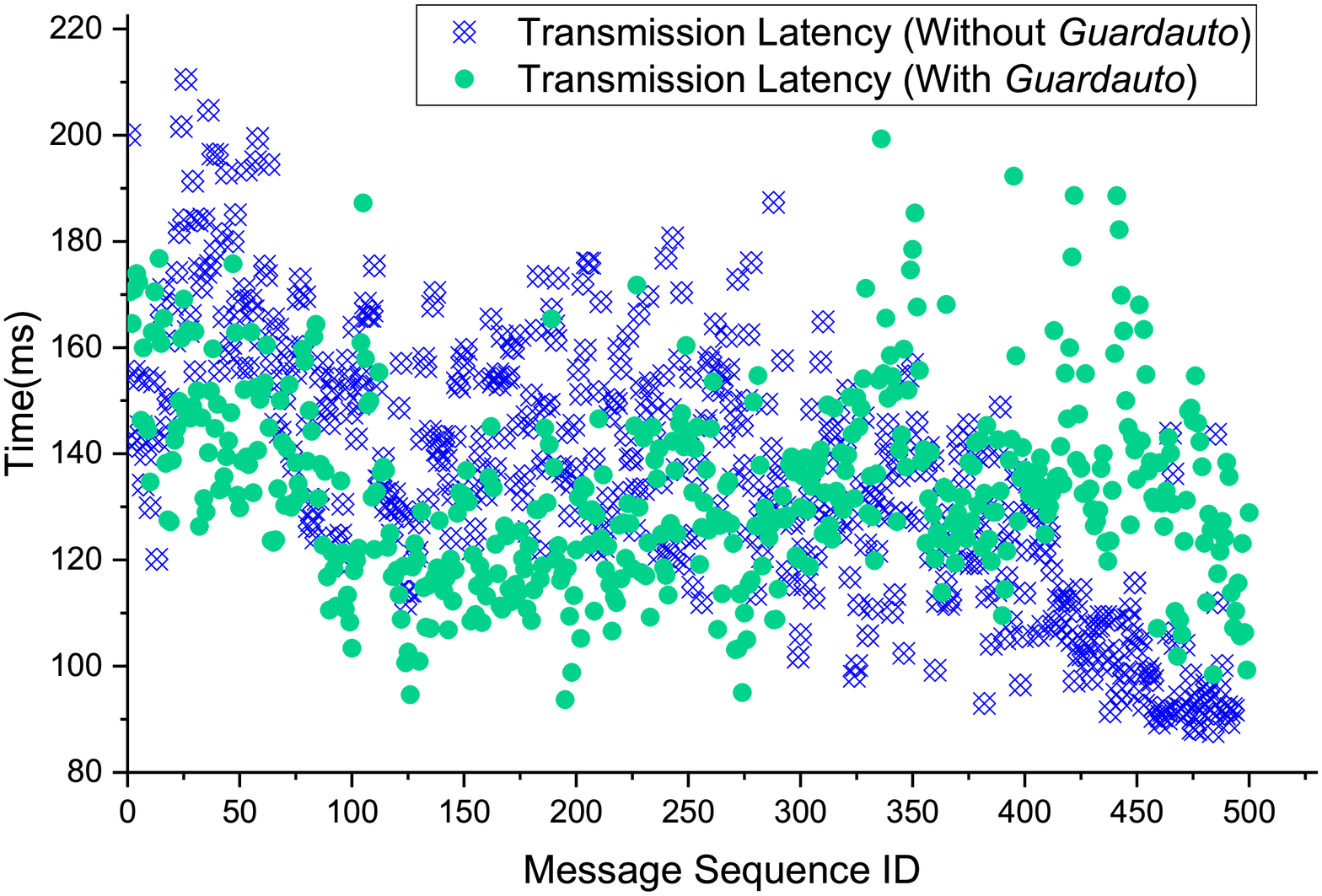}}
	\hspace{0.12cm}
	\subfigure[Points image message latency.]{
		\label{pointscomm}
		\includegraphics[width=0.28\linewidth]{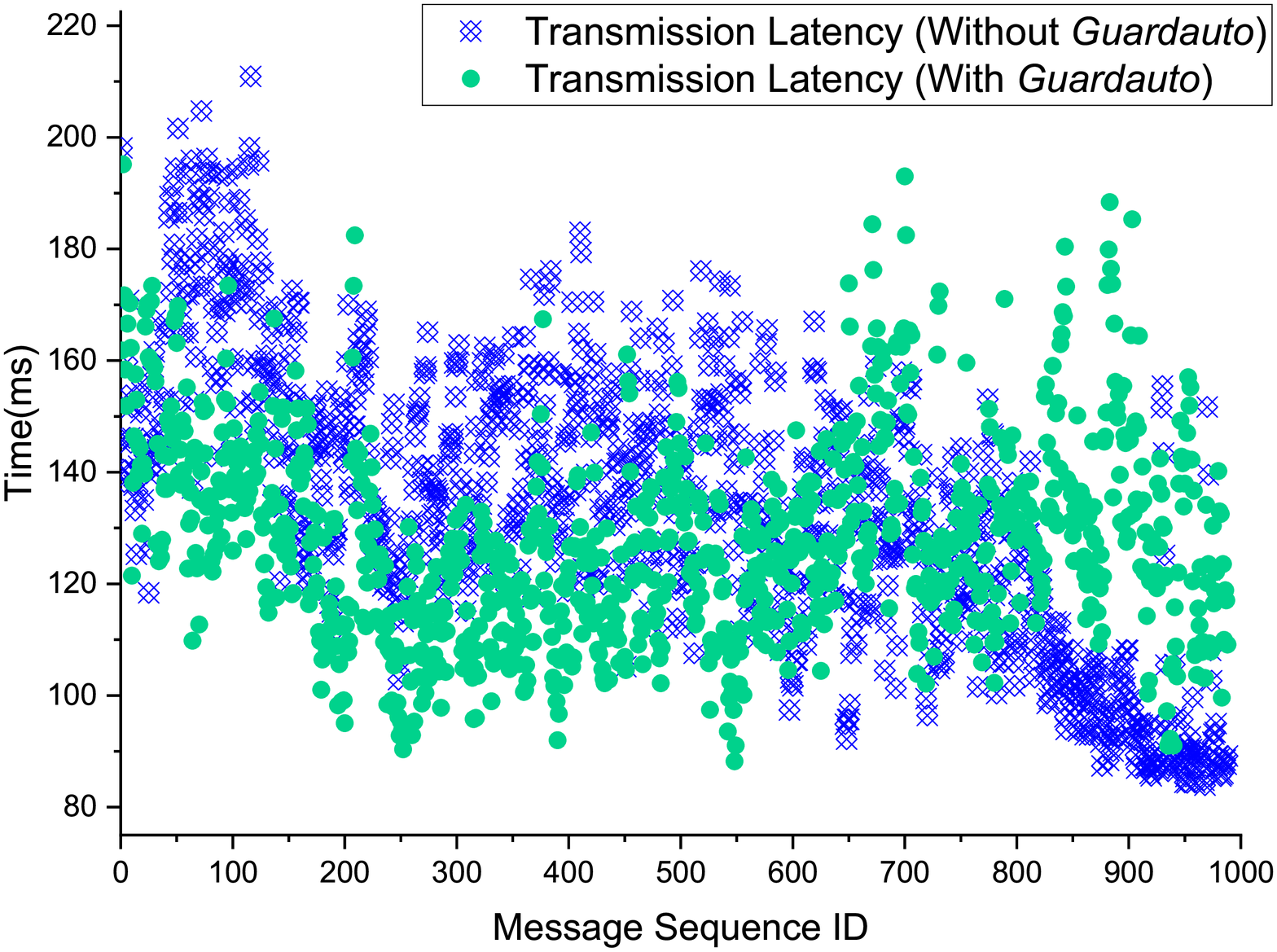}}
	\hspace{0.12cm}
	\caption{Communication latency imposed by \tool.}
	\label{comm}
	\vspace{-10pt}
\end{figure*}

The distribution of transmission latency was shown~in~Fig. \ref{comm}, where the $x$-axis denoted the message ids and the $y$-axis denoted their latency. 
Compared with~baseline results, the localization message transmission showed noticeable jitters after deploying \tool.
Because isolation increased the network complexity, the transmission in \tool jumped more hops from from one partition to another.
Moreover, as both image messages contained more bytes, the transmission cost more time than that of the localization messages, as shown in Figs.~\ref{objcomm} and~\ref{pointscomm}; and the communication jitters also introduced due to the TCP congestion control and possible re-transmission.  
On average, the localization message delay was increased by 0.23ms (from 0.38ms to 0.61ms), while for object and points image messages, the average difference was within 5\%.

%% file: SourceCode/conclusion.tex
\section{Conclusion and Future Work}\label{conclusion}
\review{2-8}{
In this paper, we propose a decentralized self-protection framework called \tool to protect self-driving systems from runtime failures or attacks.
Such a protection system decouples the autonomous driving system into components and adopts virtualization-based techniques to isolate them.
Then, \tool defends all autonomous driving system components with both local and cooperated self-protection mechanisms.
Evaluation results show that the implemented prototype meets the design goals, which can effectively detect and mitigate runtime threats.
The flexible design enables \tool great potential in securing self-driving systems.
However, the crusade of building such a system has just started, and \tool requires more efforts in especially providing more practical and innovative techniques for the failure/attack mitigation.
}

For the next step, we plan to further evaluate \tool and optimize its performance.
Second, analysis of internal interactions shall be conducted to develop a proper runtime verification approach, which can be used to reason whether the possible adaptation plan can satisfy the system requirements (e.g., reliability).
Moreover, we will extend \tool~with more fine-grained inspection approaches.
We also plan to bring the self-protection idea to secure hardware devices like ECUs.